\documentclass[lettersize,journal]{IEEEtran}
\usepackage{subcaption}

\usepackage{amsmath,amsfonts}
\usepackage{algorithm}
\usepackage{array}
\usepackage[caption=false,font=normalsize,labelfont=sf,textfont=sf]{subfig}
\usepackage{textcomp}
\usepackage{stfloats}
\usepackage{url}
\usepackage{verbatim}
\usepackage{graphicx}
\usepackage{cite}
\usepackage{balance}
\usepackage{caption}
\usepackage{amsmath,amssymb,amsfonts}
\usepackage{xcolor}
\usepackage[none]{hyphenat}
\usepackage{algpseudocode}
\usepackage[all=normal,floats, charwidths, wordspacing]{savetrees}
\usepackage{tikz} 
\newcommand{\circled}[1]{\tikz[baseline=(char.base)]{
  \node[shape=circle,draw,inner sep=1pt] (char) {\footnotesize #1};}}

\usepackage{float}
\usepackage{booktabs}    
\usepackage{tabularx} 

\begin{document}

\title{Enhancing the Security of Rollup Sequencers \\ using Decentrally Attested TEEs \\}



\author{Giovanni Maria Cristiano$^{1}$, Salvatore D'Antonio$^{2}$, Jonah Giglio$^{1}$, Giovanni Mazzeo$^{2}$, Luigi Romano$^{2}$
\thanks{$^{1}$University of Naples "Parthenope", Centro Direzionale, Isola C4, Naples, 80133, Italy. Email: \textit{first.last001@studenti.uniparthenope.it}
        }%
\thanks{$^{2}$University of Naples "Parthenope", Centro Direzionale, Isola C4, Naples, 80133, Italy. Email: \textit{first.last@uniparthenope.it}
        }%
}





\maketitle

\begin{abstract}
The growing scalability demand of public Blockchains led to the rise of Layer-2 solutions, such as \textit{Rollups}. Rollups improve transaction throughput by processing operations off-chain and posting the results on-chain. A critical component in Rollups is the \textit{Sequencer}, responsible for receiving, ordering and batching transactions before they are submitted to the Layer-1 blockchain. While essential, the centralized nature of the Sequencer makes it vulnerable to attacks, such as censorship, transaction manipulation and tampering. To enhance its security, there are solutions in the literature that shield the Sequencer inside a \textit{Trusted Execution Environment (TEE)}. However, the attestation of TEEs introduces additional centralization, which is in contrast with the core Blockchain principle. In this paper, we propose a TEE-secured Sequencer equipped with a decentralized attestation mechanism. We outline the design and implementation of our solution, covering the system architecture, TEE integration, and the decentralization of the attestation process. Additionally, we present an experimental evaluation conducted on a realistic Rollup testnet. 
\textcolor{black}{Our results show that this approach strengthens Sequencer integrity without sacrificing compatibility or deployability in existing Layer-2 architectures.}

\end{abstract}


\section{Introduction}
\label{Introduction}
\IEEEPARstart{B}{lockchain} technology has seen exponential growth since its inception, expanding its applications beyond cryptocurrency into sectors such as finance, supply chain, and healthcare. 
Despite its growing adoption and numerous potential benefits, one of the main challenges for blockchain technology is scalability \cite{scalability}. As the network grows, the number of transactions that need to be processed within a given time period becomes a bottleneck, as verifying and confirming transactions requires significant resources and time.
Many public blockchains, in fact, can only handle about twenty transactions per second (TPS) \cite{b12} \cite{b13}, a number far lower than centralized systems like traditional payment systems, which can process thousands of transactions in the same time frame \cite{b14}. To address these scalability issues, various solutions have been developed over the years, including \textit{Rollups}, also known as commit-chains \cite{b15} or validating bridges \cite{b16}.
Rollups represent one of the Layer-2 solutions, which operate on top of the main blockchain (Layer-1), aggregating multiple off-chain transactions and submitting them to the main chain as a single, compressed batch. This way, the computational load is shifted off-chain, but the final results are sent and stored on-chain. Rollups significantly reduce the amount of data and operations that the main blockchain needs to handle, thereby increasing system capacity and speed while reducing transaction fees.

\subsection{\textcolor{black}{Rationale}}

For its proper functioning, a Rollup relies on a core element: the \textit{Sequencer}. The Sequencer is responsible for ordering and bundling user transactions and then transmitting them to the main network. This makes it a crucial element for the security and reliability of the entire system. 
However, this component is vulnerable to attacks that can disrupt the system's functionality, such as transaction censorship \cite{b17}, where specific transactions are deliberately excluded, and delay attacks \cite{b18}, which postpone transaction confirmations, thus compromising system efficiency.

To address these threats, several approaches have been proposed, such as employing new security protocols to protect Rollup operations \cite{b19}, or running the entire Rollup process within a Trusted Execution Environment (TEE) \cite{b20}. This method leverages the isolation provided by TEEs to safeguard sensitive computations from external attacks and software vulnerabilities \cite{b18}, ensuring a secure and tamper-resistant execution environment.
While these solutions provide valuable improvements, they also introduce certain trade-offs. For instance, running the entire Rollup within a TEE may introduce performance overhead and increased complexity. Additionally, TEEs traditionally rely on centralized attestation processes, where a single trusted entity verifies the integrity of the environment, introducing potential vulnerabilities, such as dependence on a single point of trust.

\subsection{Proposed Solution and Contributions}
In light of these limitations, our approach focuses on isolating the Sequencer rather than the entire Rollup. This protects the ordering and bundling of transactions while avoiding unnecessary complexity and performance overhead. We also address the centralization risk of TEEs by proposing a decentralized attestation mechanism, where the TEE's quote is sent on-chain and verified via smart contracts. 
\textcolor{black}{
The main contributions of this paper are:
\begin{itemize}
\item \textbf{Targeted TEE Protection}: We propose a solution that isolates only the Sequencer within a TEE, instead of the full Rollup infrastructure. 
\item \textbf{Decentralized Attestation Mechanism}: We design and implement an on-chain verification system using smart contracts, removing the reliance on centralized attestation services.
\item \textbf{Practical Implementation}: We provide a complete implementation based on Intel SGX/TDX technologies, which share the same attestation process, and the Optimism protocol, showing feasibility in a real-world Layer-2 system.
\item \textbf{Comprehensive Evaluation}: We experimentally evaluated our solution on a locally developed Rollup testbed based on the Optimism protocol. The experimental results show that the TEE-protected Sequencer achieves a throughput of 7–8 TPS compared to 20–25 TPS in the native configuration, and confirmation latency increases from 2.5–3.5 seconds to 21–25 seconds. 
Compared to the native setup, CPU usage increases by approximately 60–80\% and memory usage increases by 30–50\%. Despite these overheads, the Sequencer gains hardware-enforced integrity and confidentiality, becoming resistant to transaction manipulation, censorship, and other host-level attacks that are possible in the unprotected setup. The decentralized attestation mechanism was also benchmarked, confirming its feasibility and manageable on-chain verification costs.
\end{itemize}
}

The quote includes runtime data from the Sequencer, selected to support precise verification. A custom policy stored on-chain guides the smart contract in ensuring that the TEE behaves as expected. Once validated, the quote is posted on-chain, providing public proof that the Sequencer ran securely. This allows anyone to verify its integrity without relying on a trusted third party.

The paper is organized as follows: Section \ref{Related_Work} reviews related work. Section \ref{Background} introduces background on Layer-1 and Layer-2 systems, the Optimism protocol, and Confidential Computing. Section \ref{Problem_Statement} defines the problem, focusing on Sequencer vulnerabilities and TEE centralization. Section \ref{Threat_MOdel} outlines our threat model. Section \ref{The_Proposed_solution} presents the proposed architecture, while Section \ref{Implementation Details} describes its implementation. Section \ref{Evaluation} evaluates the system and Section \ref{Conclusion} concludes the paper.


\section{Related Work}
\label{Related_Work}

Several solutions have been developed to improve the performance and security of Rollups, one of the most effective Layer-2 scaling methods. Two well-known examples are Arbitrum and zkSync: Arbitrum \cite{Arbitrum} uses optimistic rollups along with a bisection protocol to quickly resolve disputes, while zkSync \cite{zkSync} employs zero-knowledge proofs (ZK-SNARKs) to enhance scalability and security. StarkNet \cite{StarkNet}, another solution, uses ZK-STARKs to increase scalability without requiring a trusted setup, which is particularly useful for decentralized exchanges and smart contracts. Among these options, we chose Optimism \cite{optimismdocs} for our work, as it offers a scalable optimistic rollup solution. Its strong scalability and widespread use in various blockchain applications make it a key player in L2 scaling. 

Regarding security, various methods have been proposed to protect Layer-2 operations. For instance, L2chain \cite{L2chain} secures off-chain computations using TEEs, which prevent rollback attacks and ensure that sensitive transactions remain tamper-proof. Additionally, the Sequencer Level Security (SLS) protocol \cite{SLS}, introduced by Derka et al., adds another layer of protection by allowing the sequencer to quarantine potentially harmful transactions before they reach the blockchain. Wen et al. proposed TEERollup \cite{TEERollup},  \textcolor{black}{a non–peer-reviewed work available on arXiv,} that secures the entire rollup within TEEs, reducing on-chain verification costs and withdrawal delays by relying on TEE-backed sequencers and blockchain-anchored verification. While TEEs have been applied to secure off-chain processes or even the entire rollup, no solution to date, to the best of our knowledge, has specifically focused on securing the Sequencer itself within a TEE. This is the gap that our work seeks to address.

Efforts to decentralize TEE attestation have also emerged. DHTee \cite{DHTee} provides a decentralized infrastructure for different TEEs, using blockchain to coordinate various TEE schemes and enhance security and interoperability. This approach reduces single points of failure by decentralizing the attestation process across multiple devices.  Other approaches, such as OPERA by Chen et al.\cite{Opera}, tackle the limitations of Intel’s centralized attestation services by allowing developers to implement their own attestation processes, which enhances privacy and lowers verification times. Similarly, Zhang et al. propose JANUS \cite{Janus}, a decentralized remote attestation system that uses Physically Unclonable Functions (PUFs) in combination with blockchain to distribute trust and ensure system integrity through automated participation mechanisms.
For our work, we chose Automata \cite{Automata} to decentralize TEE attestation. Automata integrates smart contracts and provides a decentralized framework for TEE attestation, enabling trustless verification that ensures both security and decentralization.

Building on these advancements, our work is the first to propose securing the Sequencer within a TEE, while using decentralized attestation to stay aligned with blockchain’s core principles of security and decentralization.

\section{Background}
\label{Background}
In this section, we provide a background on Layer-2 blockchain solutions, the Optimism protocol, and TEEs.

\subsection{Layer-2 Blockchains} 
The three main properties of a blockchain are scalability, decentralization, and security. However, as highlighted by the so-called "scalability trilemma" \cite{trilemma1}\cite{survey1}, it is impossible to significantly improve two of these properties without compromising the third. Specifically, blockchain systems that prioritize security and decentralization, such as Bitcoin and Ethereum, often face challenges in scaling efficiently. To address this issue, various protocols have been implemented, the so-called Layer-2 solutions. \textcolor{black}{These protocols move transaction processing off-chain to improve scalability, operating on top of Layer-1 without altering its base protocol.}
Transactions are executed on Layer-2 and only periodically consolidated and recorded on the main blockchain, thereby reducing the load on Layer-1 and significantly improving throughput. One of the most promising implementations of Layer-2 solutions is \textit{Rollups}, a mechanism that groups multiple transactions into a single batch, which is then sent to the main blockchain for verification.  There are two main types of rollups \cite{rollups}:
\begin{itemize}
\item \textit{Optimistic Rollup:} Transactions are assumed valid by default but can be challenged within a time window. If a transaction is contested, an on-chain verification process is performed to ensure its validity.
\item \textit{ZK-Rollup:} use cryptographic proofs to validate and compress transactions off-chain, publishing a single proof on-chain.
\end{itemize}

\subsection{Optimism Protocol}
\begin{figure}[h]
\centering
\includegraphics[width=0.9\columnwidth]{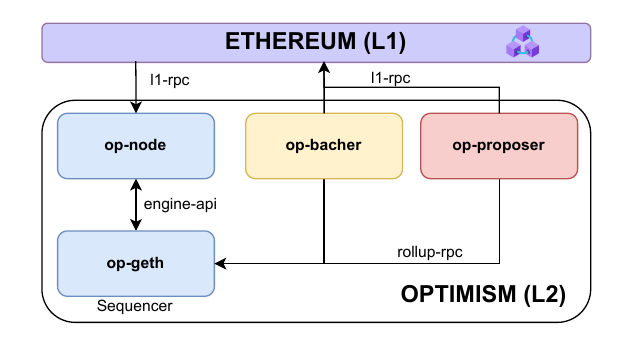}
\caption{The Optimism Protocol Architecture.}
\label{optimism_protocol}
\end{figure}

A practical example of an Optimistic Rollup implementation is the Optimism protocol \cite{optimismdocs}, summarized in Figure \ref{optimism_protocol}. Optimism is one of the most widely adopted and promising Layer-2 solutions in the context of Ethereum scalability. Since our work builds on Optimism, we briefly outline its key components. 
\\The Optimism protocol consists of several key components:

\begin{itemize}
    \item {Sequencer:} the Sequencer is a central and privileged node responsible for receiving, ordering, and executing user transactions on Layer 2. After ordering the transactions, the Sequencer creates L2 blocks and groups them into so-called \textit{sequencer batches}, which the Batcher subsequently retrieves and submits to Layer 1. The Sequencer operates off-chain but ensures that all transaction data will eventually be available on-chain. \textcolor{black}{Within the OP Stack architecture, the Sequencer node is composed of two main components:}

    \begin{itemize}
        \item op-geth (execution client): a modified version of Geth \textcolor{black}{the most widely used Ethereum execution client,} that executes transactions and updates the L2 blockchain state.
        \item op-node (consensus client): responsible for reconstructing the L2 blockchain state from data published on L1. It ensures consistency between L1 and L2 by processing L1 deposits and verifying the integrity of L2 state updates.
    \end{itemize}

    These two components interact through the Engine API, which allows op-node to request block execution from op-geth and maintain the chain’s integrity.
   
    \item {Batcher:} The Batcher (op-batcher) collects the batches created by the Sequencer, compresses and submits them to Layer-1 as calldata. Calldata is a non-executable format that allows storing transaction data efficiently on-chain. The Batcher operates off-chain and communicates with other components via RPC. Once published on L1, the batches become immutable and contribute to the final state of the Layer-2 chain.

    \item {Proposer:} The Proposer (op-proposer) is responsible for publishing \textit{state roots} to Layer-1, which represent the final state of the L2 chain after executing the transactions in a batch. While the Batcher submits raw transaction data, the Proposer commits the corresponding state, enabling Ethereum to verify the validity of L2 updates. This process is secured by a seven-day challenge window, during which any incorrect state update can be disputed through fraud proofs.
    
\end{itemize}

In Optimism, the transaction flow begins on L2, where the Sequencer collects and orders transactions into blocks. These are aggregated into batches by the Batcher and submitted to Layer-1, while the Proposer publishes the corresponding state roots.

\subsection{Trusted Execution Environment} 
TEEs create an isolated and secure area within a processor, ensuring that sensitive data and code are protected even if the main operating system or hypervisor is compromised \cite{coppolino2019survey}. Major vendors — i.e.,
AMD, ARM, Intel — proposed their TEE's implementation  \cite{Trustzone} \cite{AMD}. Among these, Intel Software Guard Extensions (SGX) \cite{SGX} and Intel Trust Domain Extensions (TDX) \cite{TDX} are prominent solutions in the field of confidential computing. 
SGX provides a process-based TEE, creating secure enclaves where selected processes are isolated from the rest of the system. TDX, on the other hand, offers a VM-based TEE, which isolates entire virtual machines (VM), providing broader protection by shielding VMs from potentially compromised host systems and hypervisors.

Intel SGX and Intel TDX both rely on the Data Center Attestation Primitives (DCAP) for remote attestation, ensuring the integrity of the execution environment. DCAP allows third-party providers, such as data centers or cloud services, to manage attestation independently, without needing continuous connection to Intel's services during runtime. In DCAP, the process begins with the generation of a cryptographic "quote" within the SGX enclave or TDX trust domain, proving its integrity. This quote is then verified using Intel’s Provisioning Certification Service (PCS), which provides the necessary certificates, including Provisioning Certification Keys (PCK), revocation lists (CRL), and information about the Trusted Computing Base (TCB). These elements ensure that the environment is secure and uncompromised.

\begin{table*}[t]
\centering
\scriptsize
\begin{tabular}{
  p{3cm}
  @{\hskip 3pt}
  p{4.5cm}
  @{\hskip 4pt}
  p{4.5cm}
  @{\hskip 4pt}
  p{4.5cm}
}
\toprule
\textbf{Asset / Goal} & \textbf{Threat / Adversary} & \textbf{Mitigation} & \textbf{Out of scope} \\
\midrule
Sequencer code/state & Transaction manipulation or censorship by host & TEE isolation; attestation & Hardware-based attacks; SGX side-channels; coordination among multiple Sequencers
 \\
Transaction ordering & Front-running sandwich attacks or injection & Enclave-managed queue; confidential execution & Non-sequencer MEV strategies; external price manipulation; off-chain data availability
 \\
Attestation quotes & Quote forgery, replay & On-chain verification & Smart-contract bugs; L1 consensus compromise \\
Communications integrity & MITM, packet tampering, DoS & Secure channels; attestation check & Network-level consensus attacks \\
\bottomrule
\end{tabular}
\normalsize
\caption{Summary of assets, threats, mitigations, and out-of-scope elements.}
\label{tab:threat_model}
\end{table*}

\section{Problem Statement}
\label{Problem_Statement}

\textcolor{black}{This section outlines the centralization risks posed by Sequencers in Rollup architectures, and the limitations of relying on a centralized TEE attestation.}

\subsection{The Problem of Sequencer Centralization}

The Sequencer, as introduced in Section III, controls the ordering and execution of transactions in Rollups. This central role gives it significant influence over transaction outcomes and makes it a potential point of failure. If the Sequencer is compromised or operated maliciously, it can exploit this position in several ways. One common category of abuse is Maximal Extractable Value (MEV). By reordering transactions, the Sequencer can perform front-running (inserting its own transaction before a user’s) or sandwich attacks (placing transactions before and after a target to profit from price changes). These strategies allow the operator to systematically extract value from honest users. The Sequencer can also engage in censorship, selectively excluding users or transaction types. This can serve personal, political, or economic purposes, such as suppressing competition or delaying time-sensitive operations. Censorship undermines trust and openness, especially in decentralized finance (DeFi) contexts.

Beyond ordering, the Sequencer can manipulate the timing and composition of transaction batches. For example, it can delay the creation of new L2 blocks or selectively include transactions to exploit race conditions or create inconsistent contract states. These actions can indirectly affect the Rollup’s behavior at Layer-1, since the Batcher depends on L2 blocks to publish data. If block production is delayed, no data reaches Layer-1, and the Rollup may revert to a fallback state containing only L1 deposits. While this prevents unsafe state progression, it breaks transaction finality and may cause denial-of-service conditions for users and applications.

To address these risks, we propose isolating the Sequencer within a Trusted Execution Environment (TEE), ensuring that its operations are executed securely and cannot be altered or influenced by the node operator or external actors.

\subsection{The Problem of TEE Centralization}

Attestation in Intel SGX and Intel TDX is essential for verifying the integrity and authenticity of applications running inside. Both of these technologies rely on the DCAP to attest the integrity of their execution environments. 
However, this process introduces centralization at two levels:
\begin{itemize}
    \item Provisioning Certification Service (PCS): A centralized service operated by Intel that provides the necessary certification data (PCK, CRL, TCB). All attestation quotes must be verified against this authority.
    \item Provisioning
    Certificate Caching Service (PCCS): The PCCS is responsible for receiving the quote and performing policy verification before validating the attestation. For example, in Azure's attestation process, the quote generated by an Intel SGX or TDX enclave is sent to a PCCS managed by Azure, where the policies are checked, and the integrity of the enclave is validated. While certificates and CRL lists can be cached locally through the PCCS, this service still relies on centralized infrastructure to receive and verify the attestation.

\end{itemize}

These components create a single point of trust. If the infrastructure is disrupted or the keys are compromised, attestation may fail or become unverifiable, leaving systems exposed. This reliance on centralized control contradicts the core principle of blockchain systems, which aim to eliminate the need for trusted intermediaries. While the PCS cannot be decentralized due to Intel’s role as hardware root of trust, our work focuses on removing the dependency on centralized PCCS services. We propose shifting policy checks and quote validation to smart contracts deployed on-chain. This decentralized attestation approach enables independent verification of Sequencer runtime data, improving resilience and aligning with the trust model of blockchain systems.

\section{\textcolor{black}{Threat Model}}
\label{Threat_MOdel}
\textcolor{black}{
This section defines the assumptions, attacker capabilities, attack vectors, and scope of the proposed TEE-secured Sequencer with decentralized attestation. A summary of covered threats and mitigations is provided in Table~\ref{tab:threat_model}.\\
\textit{Assumptions for untrust.} The host environment running the Sequencer, including the operating system, hypervisor, and container runtime, is untrusted. Network infrastructure between the Sequencer, Batcher, and Proposer is also considered untrusted, allowing adversaries to intercept, delay, or reorder messages. Availability guarantees (e.g., node uptime) are assumed but protecting against denial-of-service attacks is out of scope.\\
\textit{Assumptions for trust.} The root of trust is the TEE itself (Intel SGX), assumed to provide correct memory isolation and attestation functionality. The Layer-1 blockchain and smart contracts used for on-chain verification are assumed secure. Fraud-proof mechanisms and the 7-day challenge period are trusted to ensure data correctness. We also assume network connectivity is sufficient to maintain protocol liveness.\\
\textit{Attacker model.} We consider a privileged adversary with full control over the host infrastructure. This includes the ability to load malicious kernel modules, modify the OS, access hypervisor-level functions, or exploit vulnerabilities to escalate privileges. The attacker can attempt to:  
i) manipulate transaction ordering for economic gain (front-running, sandwiching);  
ii) selectively censor transactions;  
iii) tamper with Sequencer state updates;  
iv) forge or replay attestation quotes;  
v) compromise centralized attestation services;  
vi) perform network-level attacks such as MITM, packet injection, or targeted DoS.  
We do not consider physical attacks or speculative execution exploits (e.g., Spectre, Foreshadow), as they require conditions not typical in distributed rollup environments.\\
\textit{Threats addressed.} Our design protects the confidentiality and integrity of Sequencer execution using TEE memory isolation, preventing host-level manipulation of transaction queues. Attestation ensures verifiable execution integrity, while decentralized on-chain verification prevents forgery or reliance on centralized services. Secure channels between rollup components mitigate network tampering. Transparency in attestation metadata enables detection of censorship attempts.\\
\textit{Out of scope.} We do not address low-level hardware attacks or microarchitectural side channels, compromise of the Layer-1 consensus, bugs in verification contracts, or multi-Sequencer coordination issues. A compromise of Intel’s certificate infrastructure or TEE hardware would undermine system guarantees and is considered residual risk.
}

\section{The Proposed Solution}
\label{The_Proposed_solution}
\begin{figure}[h]
\hspace{-0.4cm}
\includegraphics[width=1\columnwidth]{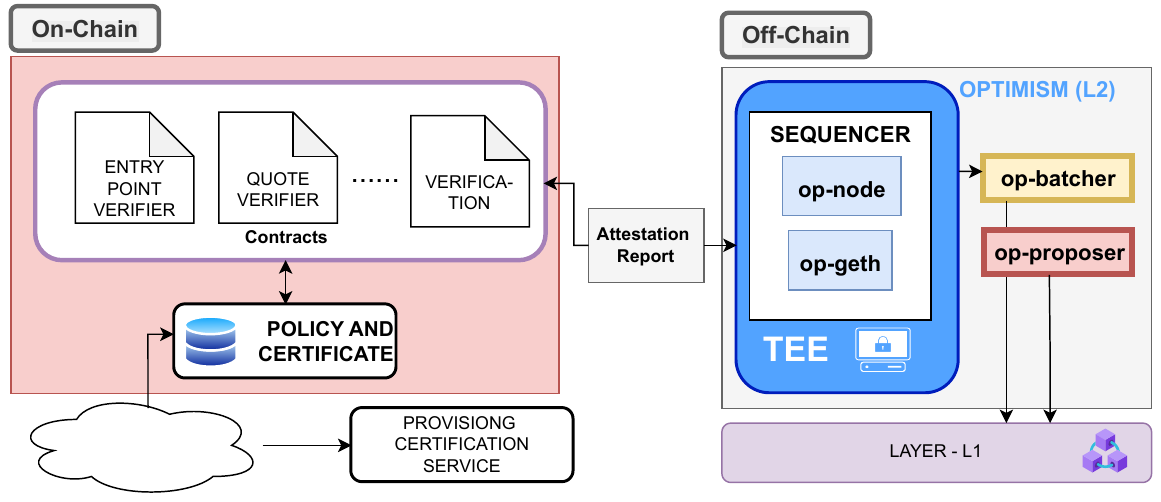}
\caption{Decentralized Attestation Architecture for Sequencer in TEE.}
\vspace{-0.5cm}
\label{Architettura}
\end{figure}

This section presents our TEE-secured Sequencer with decentralized attestation mechanism. Our approach introduces two fundamental changes to standard Rollup implementations, such as Optimism: (1) executing the Sequencer inside a TEE for isolation and integrity, and (2) implementing on-chain attestation verification through smart contracts, eliminating implicit trust in the operator and  enabling any participant to verify that a block proposed by the Sequencer was generated under verifiable and secure conditions. 
The proposed high-level architecture is shown in Figure \ref{Architettura}.

\subsection{TEE-Enabled Sequencer}

The Sequencer in our design runs within a TEE. The two core components of the Sequencer node - \textit{op-node} and \textit{op-geth} - are isolated from the host operating system and the external environment. This configuration ensures that transaction ordering, execution, and state updates are not susceptible to tampering by the node operator or by software vulnerabilities in the underlying infrastructure.
Executing the Sequencer inside a TEE guarantees that operations follow the defined protocol logic and prevents external manipulation such as transaction censorship or front-running. Unlike solutions that attempt to isolate the entire Rollup infrastructure, our approach targets only the Sequencer, which is the most critical and sensitive component. This design choice avoids unnecessary complexity and minimizes performance overhead while maintaining strong security guarantees. \textcolor{black}{While the Sequencer remains a centralized component, as in the baseline Optimism design, this is a deliberate architectural choice. Our goal is not to decentralize execution, but to significantly increase its trustworthiness by enforcing hardware-based integrity and confidentiality guarantees.}

\subsubsection{\textcolor{black}{Architectural Security Justification}}
\textcolor{black}{The proposed architectural separation between the TEE-secured Sequencer and the non-TEE components (Batcher and Proposer) is designed to maximize security while maintaining practical deployability. This separation follows Optimism's modular architecture while addressing the specific security requirements of each component.}
\textcolor{black}{The Sequencer represents the most critical attack surface in the rollup architecture. By isolating only the Sequencer within the TEE, we address the highest-impact threats while avoiding unnecessary complexity.}
\textcolor{black}{The Batcher and Proposer components operate under different security assumptions than the Sequencer: the Batcher cannot alter transaction content determined by the TEE-secured Sequencer, while the Proposer's state roots are subject to the 7-day challenge period with fraud proof protection. The security of these components relies on Layer-1's immutability and the challenge mechanism rather than requiring TEE protection.}
\textcolor{black}{In the current implementation, there is a single Sequencer instance, following Optimism's current production model. If the Sequencer is compromised, its attestation automatically expires, preventing it from publishing new blocks. A new Sequencer instance can then be initialized with a fresh TEE attestation to continue operations securely.}

\subsubsection{\textcolor{black}{Threat Containment:}} \textcolor{black}{The architectural separation ensures that compromise of the Batcher or Proposer does not affect the integrity of transaction ordering. Even if these components are compromised, the TEE-secured Sequencer continues to enforce correct transaction processing, and the Layer-1 fraud proof mechanism provides additional protection against invalid state transitions. This design creates multiple layers of security without requiring full TEE protection for all components.}

\begin{table}[H]
\centering
\renewcommand{\arraystretch}{1.5}
\resizebox{\columnwidth}{!}{ 
\begin{tabular}{|p{3cm}|p{3cm}|p{3cm}|}
\hline
\textbf{Aspect} & \textbf{Standard Optimism} & \textbf{Proposed Architecture} \\
\hline
Sequencer Execution & Runs on standard host OS & Runs inside Intel SGX enclave (via Gramine) \\
\hline
Sequencer Integrity Check & None & Attestation via SGX quote \\
\hline
Attestation Process & Not specified & On-chain via smart contract \\
\hline
Validation Model & Implicit (trust in operator) & Explicit (quote + collateral verification) \\
\hline
Block Publication & Immediately after batching & Only if attestation is valid \\
\hline
\end{tabular}
}
\caption{Comparison between Standard Optimism and the Proposed TEE-based Sequencer Architecture.}
\vspace{-0.5cm}
\label{tab:comparison}
\end{table}

\subsection{Decentralized Attestation Mechanism}
Our design decentralizes the verification of TEE attestation quotes by introducing an on-chain verification mechanism. The Sequencer generates a cryptographic quote during runtime, which includes standard attestation data (such as TEE measurement fields, signer identity, and configuration status) and Sequencer-specific metadata: block hash, block height, state root, timestamp, Sequencer nonce, and the Prover’s public key. \textcolor{black}{In our design, the Prover corresponds to the Sequencer itself running inside the TEE, which generates and signs the quote as cryptographic proof of its secure and verified execution.} This quote is transmitted to an on-chain verifier smart contract that checks its validity using an embedded policy. The verification process includes consistency checks on both TEE attestation collateral (e.g., platform certificates, revocation lists, configuration data) and protocol-level values related to the current block being produced.
A valid quote is required before block publication on Layer-1; otherwise, submission is blocked. This mechanism removes implicit trust in the node operator, ensuring that only properly attested code can alter blockchain state.

To further clarify the key differences between our approach and the standard Optimism architecture, we summarize the key aspects in the table \ref{tab:comparison}. This comparison highlights how the proposed TEE-based system improves security and decentralization, addressing limitations in the standard Optimism setup.

\subsection{\textcolor{black}{Quote Freshness and Renewal Mechanism}}
\textcolor{black}{
To ensure continuous integrity of the Sequencer, we use a dual-trigger renewal approach:
\begin{itemize}
    \item Time-based Validity Window: Each attestation quote includes a timestamp and remains valid for a predefined, configurable period stored in the Entry Point Verifier contract. The chosen validity considers security requirements, operational costs, and network finality speed. Based on our analysis in Section \ref{Implementation Details}, a 4-hour window offers a good balance between security and gas expenditure. The Sequencer periodically checks its status on-chain and, when remaining validity drops below a threshold (e.g., 20\%), it proactively submits a fresh quote to avoid interruptions. 
    \item On-demand Renewal Triggers: External parties (validators, users, or monitoring services) can request attestation renewal through the Entry Point Verifier contract by calling a requestFreshAttestation() function. To prevent denial-of-service attacks through spurious renewal requests, our system implements several protective mechanisms:
    \begin{itemize}
        \item Requester Authentication: Only whitelisted validators or governance-approved addresses can trigger renewals
        \item Rate Limiting: A minimum interval (e.g., 30 minutes) between external renewal requests is enforced
        \item Economic Disincentives: Requesters must provide a small bond (e.g., 0.1 ETH) that is slashed if the current attestation is still valid with \>50\% remaining validity
        \item Event-based Triggers: Renewal requests must provide a valid reason code (e.g., SUSPICIOUS\_ACTIVITY, CONFIG\_CHANGE) that is logged on-chain
    \end{itemize}
\end{itemize}
The smart contract enforces freshness by maintaining a mapping of sequencerAddress=[quoteHash, expirationBlock] and rejecting any block submission from Sequencers with expired attestations. This hybrid approach balances security assurance with practical considerations of gas costs and operational efficiency.
}

\begin{figure*}[t]
    \centering
    \includegraphics[width=1\linewidth]{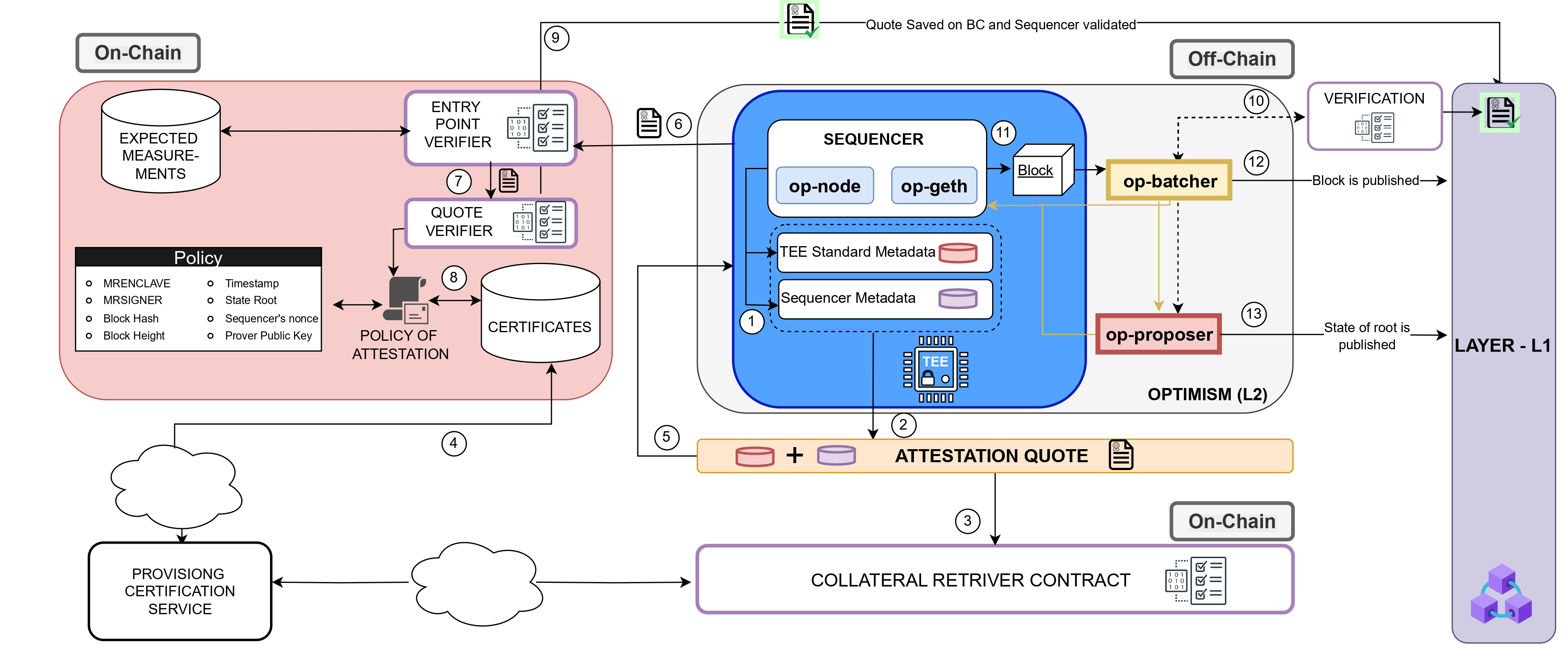}
    \caption{Execution flow of the TEE-secured Sequencer with on-chain attestation and block publication on Layer-1.}
    \label{fig:enter-label}
\end{figure*}

\subsection{Execution Flow}

The complete process, aligned with Figure~\ref{Architettura} and with the execution flow shown in Figure \ref{fig:enter-label}, follows the sequence described below. Each step refers to the smart contracts and components involved in the attestation and publishing workflow.
\\The execution process begins with the initialization of the Sequencer node, composed of \texttt{op-node} and \texttt{op-geth}, inside a TEE \circled{1}. During this phase, the enclave is measured, and integrity data is generated internally to attest to the execution environment. Once initialized, the Sequencer collects a set of block-specific metadata \circled{2}, including the block hash, block height, state root, L1 origin reference, timestamp, a unique nonce, and the Prover’s public key. These values, along with standard TEE fields (such as \texttt{MRENCLAVE} and \texttt{MRSIGNER}), are embedded into a cryptographic quote generated inside the enclave \circled{3}.
\\As part of the quote generation process, the Sequencer immediately interacts with the \textit{Collateral Retriever Contract} \circled{3}, an on-chain component responsible for retrieving attestation collateral from the \textit{Provisioning Certificate Service}. This service provides the necessary certificates (such as PCKs, CRLs, and TCB information) to validate the quote. Although decentralized in structure, this phase still depends on the provisioning infrastructure, as it is the sole authority capable of issuing the certification for the TEE platform. Once fetched, these certificates are saved on-chain and made available for subsequent verification.
After the quote is finalized, it is returned to the Sequencer and stored locally \circled{5}. Then it is submitted on chain through a transaction to the \textit{Entry Point Verifier} \circled{6}, which acts as the main gateway for the attestation infrastructure. Upon receiving the quote, the contract parses the header to determine the quote format and routes it to the appropriate \textit{Quote Verifier} contract \circled{7}, \textcolor{black}{a smart contract that validates the attestation quote and its collateral against on-chain policies.}
\\The \textit{Quote Verifier} performs all necessary validations: it checks the signature, verifies the internal structure, compares the included metadata with on-chain policies \circled{8} (such as block hash, block height, and state root), and ensures that the quote's timestamp is within an acceptable freshness window. To complete these checks, the verifier uses the attestation collateral previously retrieved and stored on-chain.
\\If the quote passes all verifications, the contract updates the on-chain attestation registry \circled{9}, marking the Sequencer as valid and authorized to publish blocks for a limited time period. This registry is then consulted by the off-chain logic of the rollup system \circled{10}, \texttt{op-batcher} and \texttt{op-proposer}, which uses it to determine whether the Sequencer is currently authorized.
If validation succeeds, the Sequencer proceeds to create a new block \circled{11} using \texttt{op-geth}, executing the ordered transactions in a secure environment. The resulting block is then passed to the \texttt{op-batcher} component, which prepares it for submission to Layer-1 \circled{12}. Before doing so, \texttt{op-batcher} checks that the Sequencer has a valid and recent attestation on record. If not, the batch is rejected and not submitted.
\\In parallel, \texttt{op-proposer} computes the corresponding state root and publishes it on Layer-1 \circled{13}. Here too, the publication is conditional on the existence of a valid quote associated with the Sequencer responsible for the block.
If, at any point, the quote is deemed invalid—due to a signature mismatch, revoked collateral, metadata inconsistency, or expired timestamp—the publication attempt is denied. In such cases, the Sequencer must regenerate a valid and fresh quote before it can continue with block proposal and submission.
\\A key challenge in the implementation was selecting which values generated by the Sequencer inside the TEE could ensure a reliable attestation system. The Block Hash and Block Height were chosen because they represent core elements of blockchain consensus, ensuring the correct sequence of blocks and their integrity. Similarly, the State Root was included to verify the global state of the blockchain, as discrepancies here could indicate a compromised Sequencer. The inclusion of the L1 Origin was crucial for verifying that the Sequencer uses the correct Layer-1 reference for Layer-2 blocks, maintaining consistency between the layers. Additionally, fields like the Timestamp and Sequencer’s Nonce were added to prevent attacks, ensuring the block's timely creation and uniqueness, thereby avoiding replay attacks.

\begin{table*}
\centering
\begin{tabular}{|c|c|c|c|}
\hline
\textbf{System Call} & \textbf{Gramine Support} & \textbf{Action Taken} & \textbf{Component} \\ \hline
\texttt{mmap()} & Supported & No changes & Memory management \\ \hline
\texttt{mremap()} & Not supported & Replaced with \texttt{mmap()} & Memory management \\ \hline
\texttt{sendmmsg()} & Not supported & Replaced with loop using \texttt{sendmsg()} & Networking \\ \hline
\texttt{ioctl()} & Partially supported & Modified file handling to avoid use & Device control \\ \hline
\texttt{inotify\_init1()} & Not supported & Removed references and replaced event handling logic & File system \\ \hline
\texttt{poll()} & Partially supported & Adjusted code to use supported alternatives & Networking \\ \hline
\texttt{epoll\_wait()} & Supported & No changes & Event polling \\ \hline
\texttt{open()} & Partially supported & Rewrote file operations to avoid unsupported paths & File handling \\ \hline
\texttt{fork()} & Supported & No changes & Process management \\ \hline
\texttt{sched\_yield()} & Supported & No changes & Thread management \\ \hline
\end{tabular}
\caption{Changes made on the Sequencer for its TEE-fication. System calls in the table refer to the \texttt{op-geth} component.}
\label{table:1}
\end{table*}
\section{Implementation Details}
\label{Implementation Details}

In our implementation, we focused on Intel SGX as the Trusted Execution Environment for securing the Sequencer. Although both SGX and TDX share the same attestation model and can be used in similar scenarios, SGX was selected for two main reasons. First, SGX provides a smaller Trusted Computing Base (TCB) compared to TDX, offering a more minimal and auditable environment, which reduces the potential attack surface. Second, despite its stricter isolation, SGX typically results in higher overhead in terms of performance, as shown by recent studies \cite{PerformanceAnalysis1}. By selecting SGX, we intentionally evaluated our system in a more constrained setting, representing a worst-case scenario from a performance perspective. This choice allowed us to better assess the trade-offs between security and efficiency, and to verify the practicality of our decentralized attestation mechanism under more demanding conditions.

We based the Rollup implementation on the Optimism protocol, leveraging its Layer-2 scaling to enhance Ethereum’s performance. For decentralized attestation, we adopted elements of Automata's architecture \cite{Automata}, which uses a blockchain-based network to verify TEE attestations, distributing verification responsibility across multiple parties and improving transparency. Automata supports SGX enclaves and provides on-chain infrastructure for quote and collateral verification. The following sections provide details on how we integrated these components.

\subsection{The TEE-secured Sequencer}
To securely run the Sequencer, we used Gramine \cite{Gramine}, a library OS that allows unmodified (or minimally modified) applications to run inside a TEE such as Intel SGX. The critical components we moved inside Gramine were the "op-node" and "op-geth," which togheter form the Sequencer node. Running these components inside Gramine required several code changes, as Gramine, while theoretically compatible with the Go programming language, presented limitations, especially regarding system calls used by these components. 
Some libraries --- particularly those related to file handling or network interactions --- were not fully supported by Gramine. This forced us to rewrite parts of the code or remove non-essential features. Additionally, several system calls used by the Optimism components were not compatible with the Gramine runtime environment. 
Our approach involved two strategies: replacing unsupported system calls with compatible alternatives and removing non-essential system calls.

\begin{itemize}
    \item Replacing Incompatible System Calls: in cases where alternative system calls were available, we replaced the unsupported ones while preserving the original functionality. For instance, the memory management system call \textit{mremap()}, which is not supported by Gramine, was replaced with \textit{mmap()}. This allowed us to maintain dynamic memory resizing functions using a compatible system call. Similarly, for networking, the system call \textit{sendmmsg()}, which is used for batching messages, was not supported. We modified the networking layer to use \textit{sendmsg()} in a loop to handle multiple messages, ensuring that the logic remained intact while adhering to Gramine's supported call set.
    \item Removing Non-Essential System Calls: in some cases, we determined that certain system calls were not critical for the core functionality of the Sequencer. These calls were eliminated to prevent failures in the Gramine environment. For example, \textit{inotify\_init1()}, part of the fsnotify library used for monitoring file system events, was removed along with its references in files such as backend\_inotify.go and go.mod, as it did not impact the primary functions of the Sequencer. Also some shared memory system call, like \textit{shmeg()} and \textit{shmat} are not supported by Gramine. These were primarily used in inter-process communication (IPC) mechanisms that could be re-engineered to use other techniques like file-based communication, thus simplifying the architecture while eliminating unsupported calls.
\end{itemize}

A summary of these system calls and their status is provided in Table \ref{table:1}, which outlines specific actions taken to resolve incompatibilities for \textit{op-geth}, one of the core components of the Sequencer. To ensure the successful execution of the Sequencer inside the TEE, we also created an appropriate Gramine manifest file, detailing the required resources and setting memory limits for the enclave. 

\subsection{The Decentralized Attestation of the Sequencer}

The implementation of the decentralized attestation mechanism was built using Solidity 0.8.0, which was used to develop the smart contracts responsible for managing the on-chain attestation process. This version of Solidity was chosen for its enhanced safety features, such as automatic overflow checks, which increase the security and robustness of the smart contracts. Our system integrates key libraries from Automata, specifically Automata-dcap-attestation\cite{Automata1} and Automata-dcap-qpl\cite{Automata2}, to handle the decentralized verification of the quotes and collaterals. Automata-dcap-attestation was used to manage the submission and verification of the cryptographic quote generated by the Sequencer's TEE. This library allowed us to streamline the attestation flow by automating the parsing and validation of TEE quotes on-chain. The core contract, AutomataDcapAttestation.sol, was integrated into our architecture to ensure that every quote submitted to the blockchain undergoes thorough verification.

In terms of customization, we adapted the Automata-dcap-attestation library to verify additional fields specific to the Sequencer, such as the Block Hash, Block Height, and State Root. These fields were added to ensure the consistency and security of the rollup protocol. For example, the \textit{verifyAndAttestOnChain} function was extended to check not only the standard SGX quote fields but also Sequencer-specific values, ensuring that the block's integrity is maintained before it is added to the blockchain.

To further secure the validation process, we employed Automata-dcap-qpl, which handles the management and verification of collaterals, including PCK certificates, CRLs, and TCB status. This library enables the on-chain smart contracts to communicate with Intel’s PCS, ensuring that the required certificates are validated without relying on centralized services like the traditional PCCS. The \textit{verifyCollaterals} function was key to ensuring that the collaterals provided within the quote were valid and had not been revoked.

In cases where discrepancies were found—such as a mismatch in the Block Hash or an expired PCK certificate—the smart contract was designed to automatically reject the quote and trigger an alert. This action prevents any compromised block from being published on-chain, maintaining the integrity of the rollup protocol. By integrating and customizing these libraries, we were able to create a robust and decentralized attestation system that operates autonomously on-chain, providing both transparency and security for the Sequencer’s operations.

\section{Evaluation}
\label{Evaluation}
The experimental evaluation was designed to assess the performance of the Sequencer in two configurations: (i) an unprotected execution, where the Sequencer operates in a standard environment without additional security measures, and (ii) a TEE-protected execution, where the Sequencer runs within an Intel SGX enclave using Gramine. In this way, we quantify the computational overhead introduced by the TEE.

\subsection{Testbed}

The experimental setup was conducted on a Microsoft Azure Standard\_DC4s\_v3 virtual machine equipped with an Intel® Xeon® Platinum 8370C CPU @ 2.80GHz (4 vCPUs), 32 GB of RAM, 256 GB of storage and Ubuntu 22.04.4 LTS as the operating system. The system supports Intel SGX, enabling secure enclave-based computations. To evaluate the performance of the Sequencer, we deployed a custom L2 Rollup testnet based on the Optimism protocol [citation to Optimism documentation], with Sepolia as the underlying Layer-1 blockchain and Alchemy as the RPC provider. 
\\To ensure the reproducibility of results, the following versions of software components were used. The Optimism stack consisted of op-node v.1.9.4, op-proposer v.1.9.4, op-batcher v.1.9.4, op-geth  (which is based on geth v1.14.13) and go v1.23. For the TEE execution, Gramine v1.7 was employed as the runtime environment. For the TEE-based execution, Gramine v1.7 was used as the enclave runtime environment, along with the Intel SGX SDK, configured to support Data Center Attestation Primitives (DCAP) on the Azure platform.

\subsection{Design of Experiments}

The benchmarking of the Sequencer's transaction processing capabilities was carried out using a dedicated script, designed to automate the generation and submission of transactions to the testnet. The benchmark was structured to evaluate different configurations of transaction volume and payload size. Specifically, the number of transactions per test was set to 10, 50, 100, 200, 300, 500, and 1000, while the payload sizes ranged from 100B to 5000B. These values were selected based on real-world Ethereum transaction characteristics. Simple ETH transfers typically range between 200B and 250B, whereas transactions involving smart contract execution may reach sizes of 2KB to 10KB. Payloads exceeding 50KB are uncommon in practical scenarios, making the chosen range representative of real-world blockchain workloads.

The decision to limit the number of transactions per test was primarily driven by economic and computational constraints. While the tests were conducted on the Sepolia testnet, transactions still incurred gas costs, and mining the necessary testnet ETH required significant time. Given these constraints, a higher transaction count was deemed impractical. Similarly, the number of concurrent workers executing transactions was set to 100, as exceeding this threshold would have led to excessive computational overhead on the available hardware, without providing meaningful performance insights.
The experimental setup and test design are summarized in Table \ref{tab:exp_setup_design}.

\begin{table}[ht]
\centering
\begin{tabular}{@{}l p{6cm}@{}}
\toprule
\textbf{Parameter} & \textbf{Value / Description} \\
\midrule
\multicolumn{2}{@{}l}{\textbf{Hardware and Software Setup}} \\
\midrule
Cloud Provider & Microsoft Azure (Standard DC4s v3) \\
CPU           & Intel Xeon Platinum 8370C @ 2.80\,GHz (4 vCPUs) \\
RAM           & 32\,GB \\
Storage       & 256\,GB \\
Operating System & Ubuntu 22.04.4 LTS \\
TEE Support   & Intel SGX (Gramine\,v1.7, SGX SDK with DCAP) \\
Optimism Stack & op-node v.1.9.4, op-proposer v.1.9.4, op-batcher v.1.9.4, op-geth (based on geth v1.14.13), go v1.23 \\
\midrule
\multicolumn{2}{@{}l}{\textbf{Benchmark Configuration}} \\
\midrule
Underlying L1 & Sepolia testnet (Alchemy as RPC provider) \\
Transaction Counts & 10, 50, 100, 200, 300, 500, 1000 \\
Payload Sizes  & 100\,B, 300\,B, 500\,B, 1000\,B, 2000\,B, 3000\,B, 5000\,B \\
Concurrent Workers & 100 \\
Rationale      & Limited transaction counts and concurrency \\
               & due to gas costs and hardware constraints \\
\bottomrule
\end{tabular}
\caption{Experimental setup and test design summary.}
\label{tab:exp_setup_design}
\end{table}

To comprehensively analyze system behavior under load, both Sequencer performance and system resource utilization were continuously monitored. Four primary metrics were monitored and evaluated during these experiments:

\begin{itemize}

    \item \textbf{Startup Time:} Measures the initial overhead introduced by executing the Sequencer within a TEE, specifically focusing on the average startup time (the time required to fully initialize and launch the Sequencer node, from enclave creation to when the Sequencer becomes operational) and the associated CPU usage during initialization.
    \item \textbf{Latency:} Calculated as the average time elapsed between transaction submission by a client and the acknowledgment of transaction receipt within the Sequencer's mempool. This metric directly affects the perceived responsiveness of the Sequencer, critical in scenarios where timely processing and quick acknowledgment are essential.
    \item \textbf{System throughput:} Defined as the average number of transactions successfully processed per second (TPS). This metric highlights the overall transaction processing capability of the Sequencer under different workloads, allowing us to identify performance impacts caused by executing within the SGX environment.
    \item \textbf{Resource consumption:} Evaluates the overall resource utilization (CPU and RAM) across the entire Rollup system during transaction processing. The measurement includes resource usage by both the TEE-enabled Sequencer (running within the SGX enclave) and the standard Sequencer configuration. 
\end{itemize}

\subsection{Performance Analysis}

\subsubsection{Startup Performance Analysis} Table \ref{tab:performance_metrics_2} summarizes the results of the startup performance analysis, highlighting the overhead introduced when initializing the Sequencer within an SGX enclave compared to a standard, unprotected environment.

As clearly illustrated, executing the Sequencer within an SGX enclave significantly increases the average startup time, rising from less than one second in the standard execution scenario to approximately 8 seconds within the TEE. Similarly, CPU usage during startup shows a noticeable increase, from roughly 5\% in the unprotected scenario to around 24\% when utilizing Intel SGX. This increase is due primarily to enclave initialization procedures, cryptographic setup, and secure memory allocations inherent in SGX environments.
While these overheads impact system responsiveness at initialization, they represent a trade-off justified by the enhanced security and isolation offered by the TEE.

\begin{table}[ht]
\centering
\begin{tabular}{lcc}
\hline
\textbf{Metric} & \textbf{With TEE} & \textbf{Without TEE} \\
\hline
Average Startup Time(s) & 8s & 0.9s \\
Average CPU Usage (\%)  & 24\% & 5\% \\
Peak Memory Usage (MB)  & 82 MB & 61 MB \\
Average Context Switches (count)  & 6000 & 1200 \\
\hline
\end{tabular}
\caption{Startup Performance Metrics}
\label{tab:performance_metrics_2}
\end{table}

\subsubsection{Latency Analysis} 
\begin{figure}
    \vspace{-0.5cm}   
    \begin{subfigure}{0.50\textwidth}
        \centering
        \includegraphics[width=\linewidth]{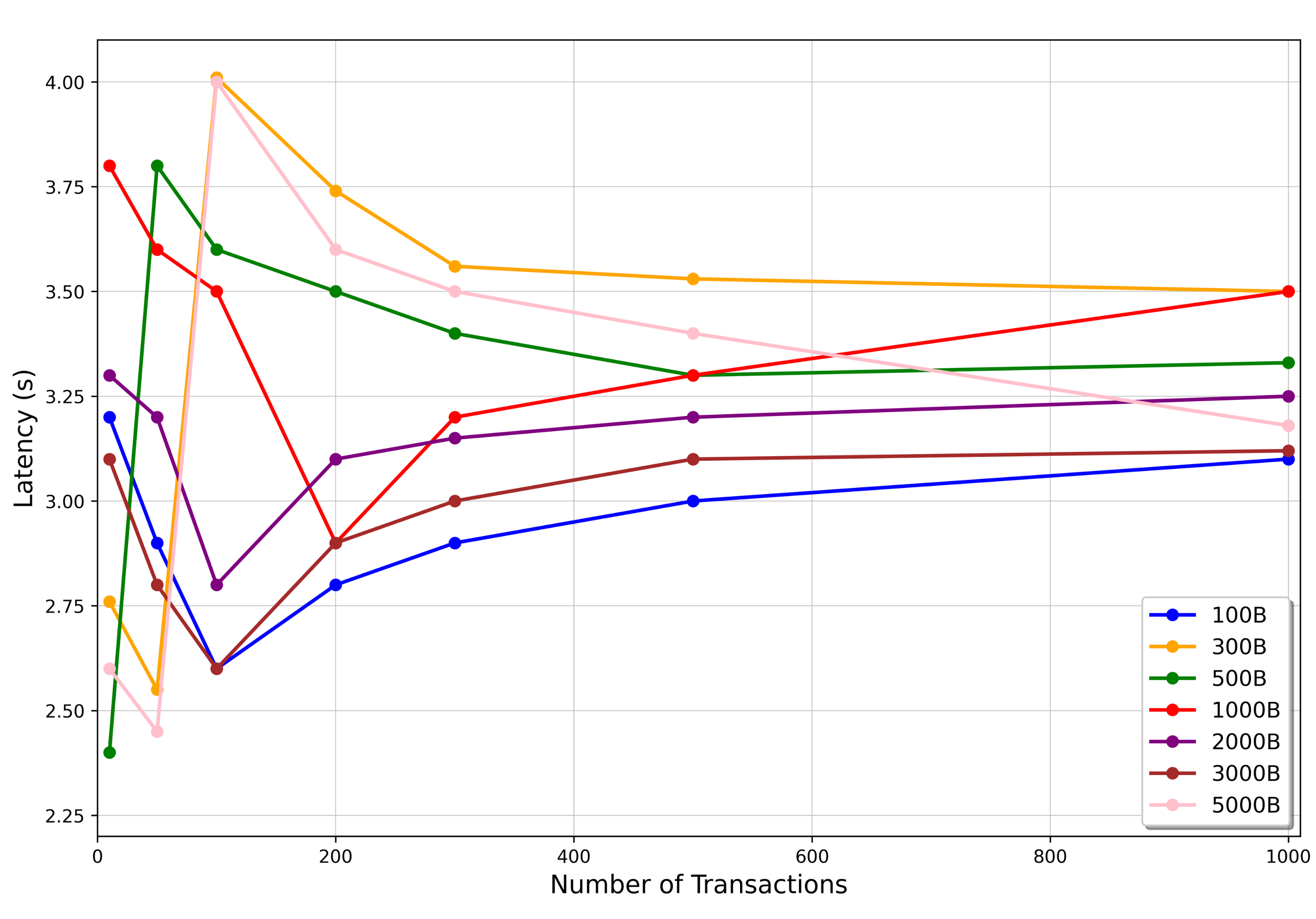}
        \caption{Latency without TEE}
        \label{fig:latency_no_tee}
    \end{subfigure}
    \hfill
    \begin{subfigure}{0.50\textwidth}
        \centering
        \includegraphics[width=\linewidth]{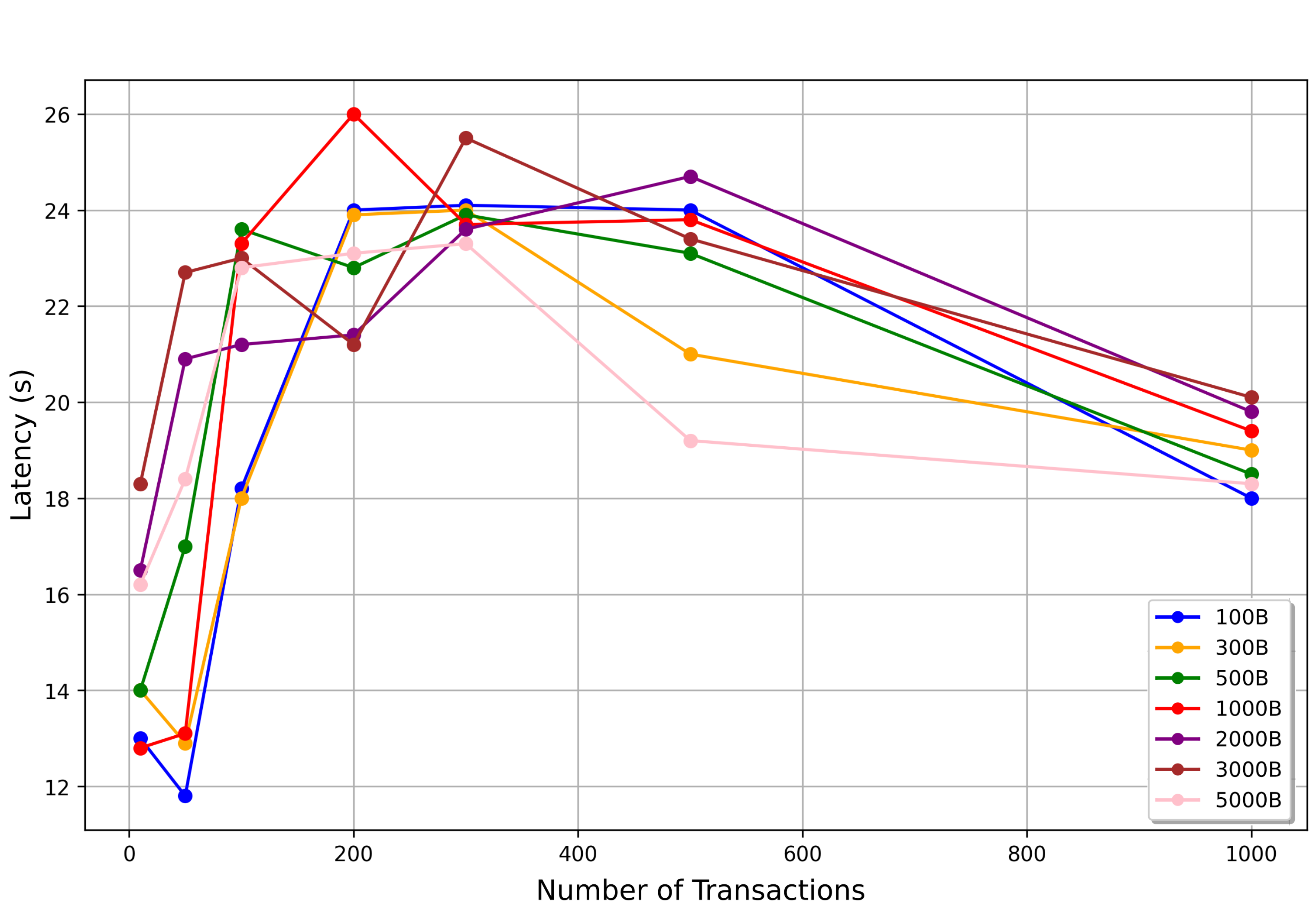}
        \caption{Latency with TEE}
        \label{fig:latency_tee}
    \end{subfigure}
    \vspace{-0.5cm}
    \caption{Latency of the Optimism protocol with the Sequencer executed (a) outside a TEE and (b) inside a TEE.}
    \label{fig:benchmark_latency}
    \vspace{-0.5cm}
\end{figure}

Transaction latency is a critical metric for evaluating the responsiveness of the Sequencer. Figure \ref{fig:benchmark_latency} shows a clear difference between the two execution environments, but also reveals common trends in both scenarios.
In the unprotected setup (without TEE), latency initially presents considerable variability, particularly for low numbers of transactions. The latency values fluctuate, ranging from approximately 2.5 to nearly 3.75 seconds. However, as the transaction volume increases, a clear stabilizing trend emerges: latency consistently settles within a narrower and lower range, approximately between 3.0 and 3.5 seconds, independent of payload size. 
Conversely, when the Sequencer is executed within a TEE (Intel SGX), overall latency significantly increases, with values starting around 21-23 seconds and experiencing fluctuations that exceed 25 seconds for certain payloads and transaction volumes. Despite the evident higher latency compared to the unprotected scenario, a similar positive trend emerges: latency values initially fluctuate but tend to gradually stabilize (around 23-24 seconds) as the transaction count increases, eventually reducing slightly and becoming more consistent. In both scenarios, this behavior can be attributed to network optimization mechanisms or improved efficiency in batching and processing transactions under higher load conditions. 

A portion of the elevated latency can be attributed to the execution environment itself: Intel SGX imposes additional overhead due to enclave creation, enclave entry/exit calls, and secure memory management \cite{PerformanceAnalysis1}. Another important factor influencing these results is the experimental setup, which runs an entire Optimism stack locally within a single node. In a real-world deployment with optimized infrastructure (such as distributed RPC nodes and optimized network connectivity), it is reasonable to expect that this latency would be reduced significantly, although it would still remain higher compared to non-TEE solutions.

\subsubsection{Throughput Analysis}

\begin{figure}[h]
    \vspace{-0.5cm}   
    \begin{subfigure}{0.50\textwidth}
        \centering
        \includegraphics[width=\linewidth]{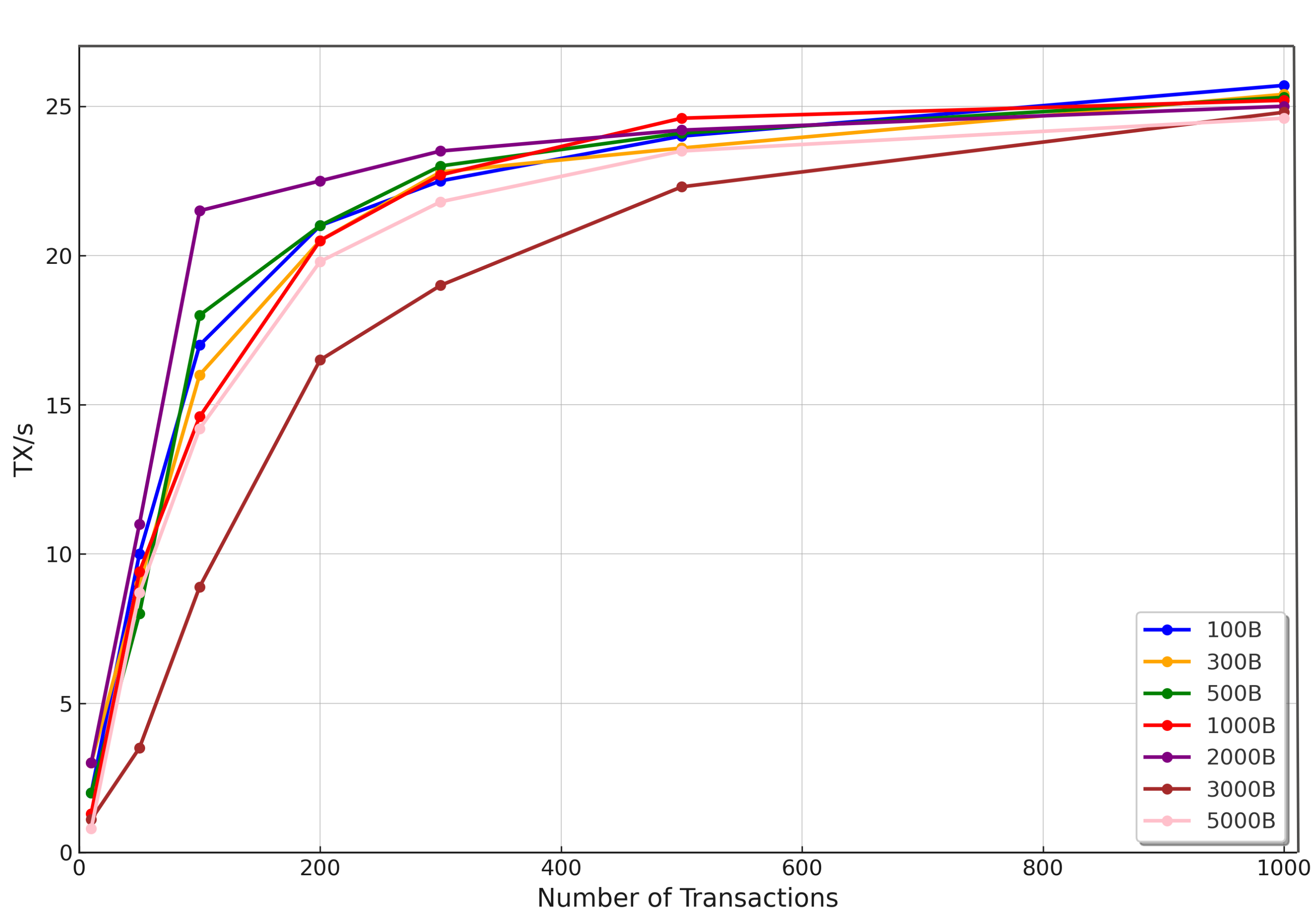}
        \caption{Throughput without TEE}
        \label{fig:latency_no_tee}
    \end{subfigure}
    \hfill
    \begin{subfigure}{0.50\textwidth}
        \centering
        \includegraphics[width=\linewidth]{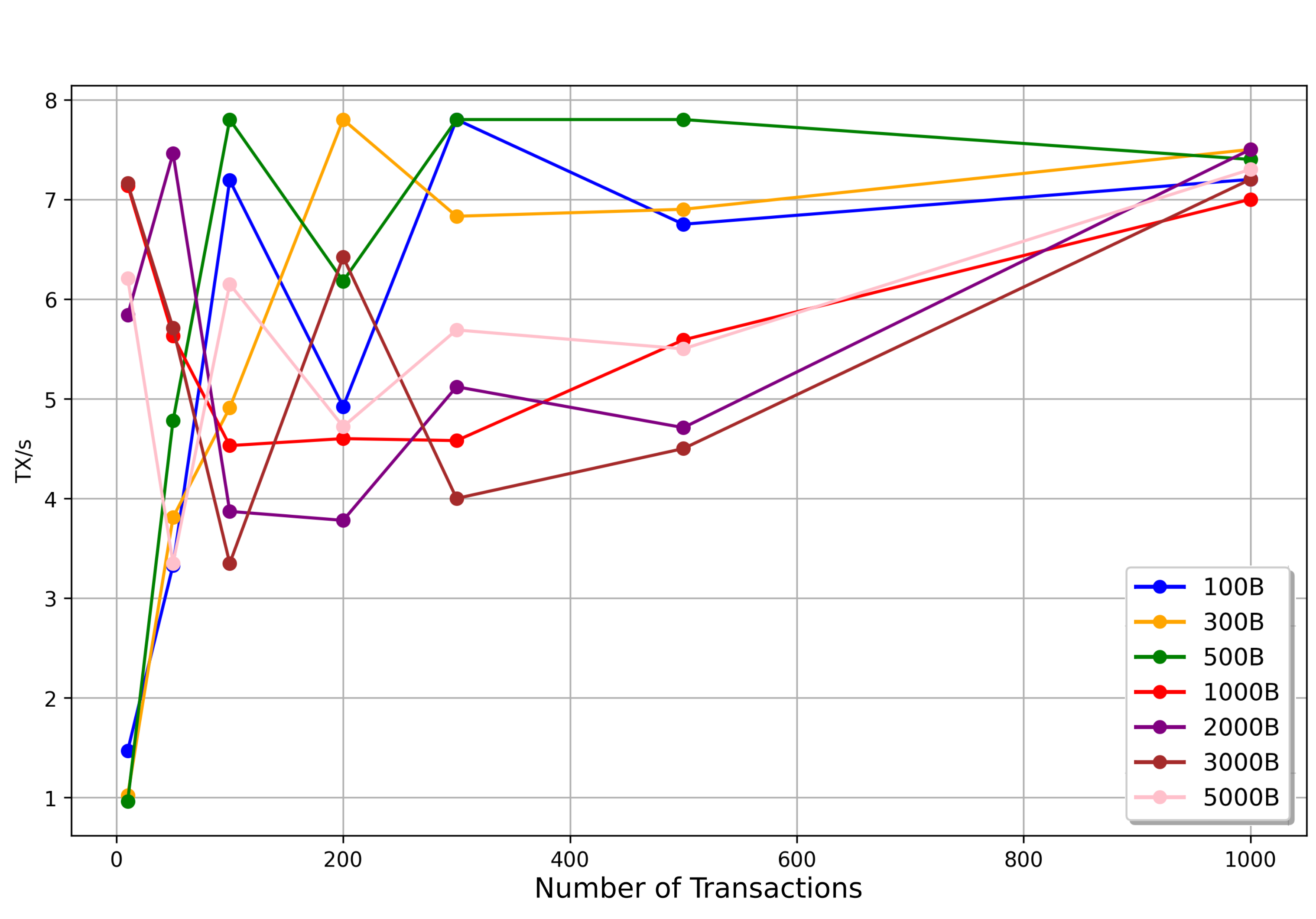}
        \caption{Throughput with TEE}
        \label{fig:latency_tee}
    \end{subfigure}
    \vspace{-0.5cm}
    \caption{Throughput of the Optimism protocol with the Sequencer executed (a) outside a TEE and (b) inside a TEE.}
    \label{fig:benchmark_throughput}
\end{figure}

Figure \ref{fig:benchmark_throughput} shows the throughput (TPS) of the Sequencer in both configurations. Without TEE, throughput quickly stabilizes around approximately 25 transactions per second, for all payload sizes as transaction count increases. This result indicates a consistent performance of the Sequencer in standard conditions, suggesting minimal resource contention even under higher workloads. 

In contrast, executing the Sequencer inside the TEE significantly affects throughput, reducing it to approximately 7-8 TPS at best. This performance is about three times lower compared to the unprotected execution. Moreover, the TEE-protected Sequencer exhibits notable variability in throughput as the transaction number and payload size change. Specifically, throughput appears unstable and fluctuating under different load conditions, with smaller payloads (100-500 bytes) generally showing slightly better performance compared to larger payloads (2000-5000 bytes).
This decline highlights the computational overhead introduced by Intel SGX, primarily due to encryption overhead, context switching, and enclave memory management, which grow more intensive as payload size increases.
In addition to the differences between the TEE-protected and unprotected configurations, a noticeable positive trend emerges in both scenarios. Initially, when processing a lower number of transactions (e.g., 10 or 50), the Sequencer achieves relatively lower throughput, both with and without TEE. As the transaction load increases, the throughput gradually improves, eventually stabilizing at higher transaction counts (around 600-1000 transactions). This trend can be attributed to the nature of transaction batching within the Sequencer. With smaller workloads, the overhead associated with transaction handling and system setup (such as networking latency, memory management, and context-switching operations) proportionally impacts throughput more heavily, resulting in lower observed TPS. As the number of transactions increases, the overhead associated with batching and submission becomes amortized more efficiently between transactions, leading to improved throughput.

\begin{figure}[h]
    \vspace{-0.5cm}   
    \begin{subfigure}{0.50\textwidth}
        \centering
        \includegraphics[width=\linewidth]{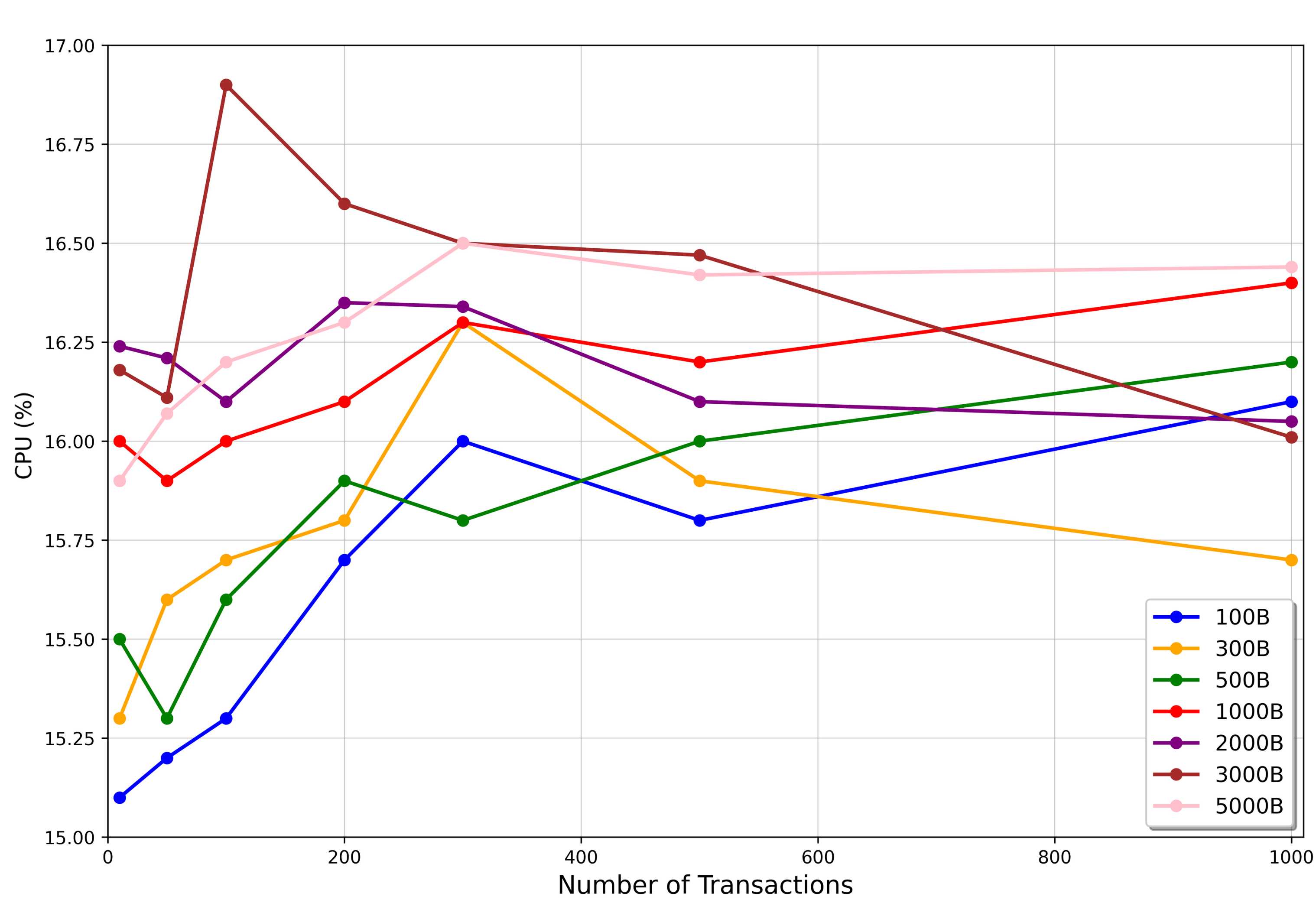}
        \caption{CPU usage without TEE}
        \label{fig:cpu_no_tee}
    \end{subfigure}
    \hfill
    \begin{subfigure}{0.50\textwidth}
        \centering
        \includegraphics[width=\linewidth]{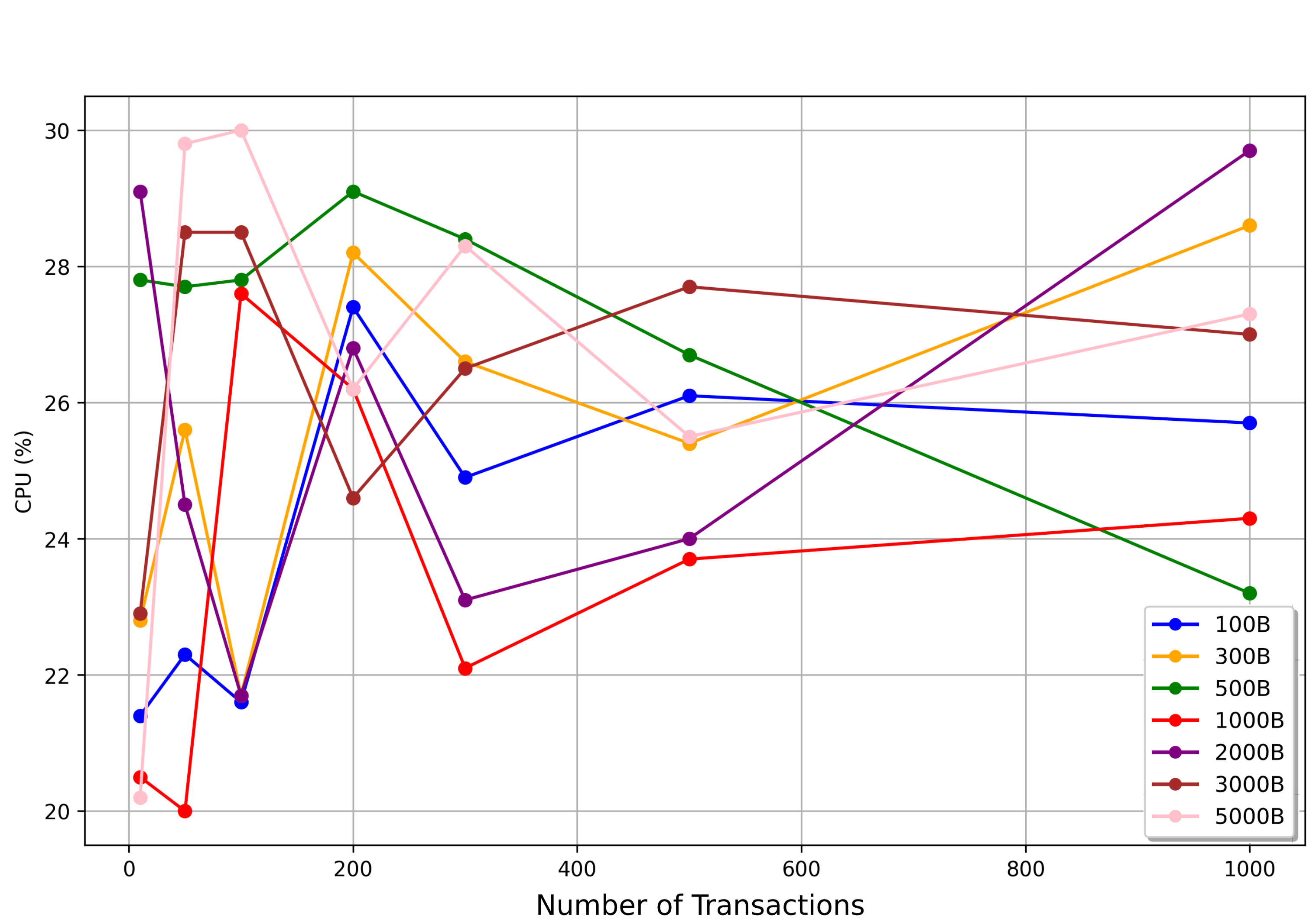}
        \caption{CPU usage with TEE}
        \label{fig:latency_tee}
    \end{subfigure}
    \vspace{-0.5cm}
    \caption{Comparison of CPU consumption when executing the Sequencer (a) outside a TEE and (b) inside a TEE.}
    \label{fig:benchmark_cpu}
\end{figure}

\subsubsection{CPU and Memory Resource Consumption}

Resource utilization, specifically CPU and RAM usage, provides information on the computational demands introduced by running the Sequencer within a TEE compared to the standard execution scenario. The experimental evaluation revealed significant differences between these two setups.
\\In the unprotected configuration, CPU consumption remains relatively low and consistent across varying workloads. As depicted in the provided Figure \ref{fig:benchmark_cpu}, CPU utilization fluctuates modestly in a range of approximately $15.7\%$ to $16.3\%$. These results highlight efficient CPU usage, indicating minimal overhead from transaction processing tasks in the non-TEE environment. In contrast, executing the Sequencer inside the Intel SGX enclave notably increases CPU usage. In this configuration, CPU consumption approximately doubles, ranging between 24\% and 30\%. This substantial increase aligns well with previous studies and can be attributed to additional computational demands from SGX-specific tasks. These operations cumulatively intensify the CPU workload, significantly impacting the overall usage of resources.
Additionally, CPU utilization in the TEE scenario appears highly irregular, especially for smaller transaction batches. This variability may be due to initial overhead related to enclave initialization, cryptographic setup operations, and the limited ability of the enclave environment to optimize batch processing efficiently at lower transaction volumes. However, as the transaction count increases, CPU utilization tends to stabilize, suggesting a better distribution and amortization of SGX overheads at higher workloads.

For what concerns the memory usage, the TEE execution scenario similarly leads to increased resource consumption compared to the standard execution. In particular, RAM usage increases significantly due to SGX requirements for memory isolation, encrypted paging, and secure page management, consistent with findings in the previous literature \cite{coppolino2019survey}. The memory overhead introduced by the SGX enclave intensifies under heavier workload conditions and larger transaction payloads, reflecting the enclave's requirement to manage and protect transaction data securely.

\begin{table}
\centering
\scalebox{0.85}{
\begin{tabular}{@{}lccp{3.2cm}@{}}
\toprule
\textbf{Metric} & \textbf{Unprotected} & \textbf{TEE} & \textbf{Notes} \\
\midrule

\textbf{Startup time}      
  & 0.9\,s      
  & 8\,s       
  & Enclave initialization overhead \\

\textbf{Context Switches} 
    & 1200  
    & 6000 
    & Additional overhead from enclave switching \\
  
\textbf{Latency}      
  & 2.5--3.5\,s      
  & 21--25+\,s       
  & SGX enclave overhead \\

\textbf{Throughput}   
  & \(\sim\)25 TPS   
  & 7--8 TPS         
  & Impact of encryption \& switching \\

\textbf{CPU Usage}    
  & 15.7--16.3\%     
  & 24--30\%         
  & Higher at lower transaction volumes \\

\textbf{Memory Usage} 
  & 19-20\% 
  & 20-30\% 
  & Due to memory isolation/encryption \\
\bottomrule
\end{tabular}}
\caption{Summary of performance metrics for unprotected vs.\ TEE-based execution.}
\label{tab:performance_summary}
\end{table}

These results can be generalized as follows (Table \ref{tab:performance_summary}): 
\begin{itemize}
    \item Latency: Running the Sequencer within a TEE increases latency from about 2.5–3.5 seconds to over 25 seconds. This gap limits feasibility in scenarios requiring fast transaction finality, such as decentralized exchanges or payment systems. However, it can still be acceptable in settings where security and data integrity take priority, for example in asset custody or high-value settlements.
    \item Throughput: The unprotected configuration stabilizes at about 25 TPS, whereas TEE execution falls to around 7–8 TPS. This decline is due to enclave overhead (e.g., memory encryption and context switching). While the drop is significant, it may be justified in use cases focusing on data integrity or censorship resistance over raw speed.
    \item CPU and Memory Usage: Without TEE, CPU usage hovers around 15–16\%. With TEE, it can exceed $30\%$, particularly at lower transaction volumes where enclave initialization and cryptographic tasks have a greater impact. Memory usage also rises substantially when running inside the enclave, reflecting the cost of memory isolation and encryption. At higher transaction rates, these overheads become more distributed and resource usage appears more stable.
\end{itemize}

\begin{figure}[h]
    \begin{subfigure}{0.50\textwidth}
        \centering
        \includegraphics[width=\linewidth]{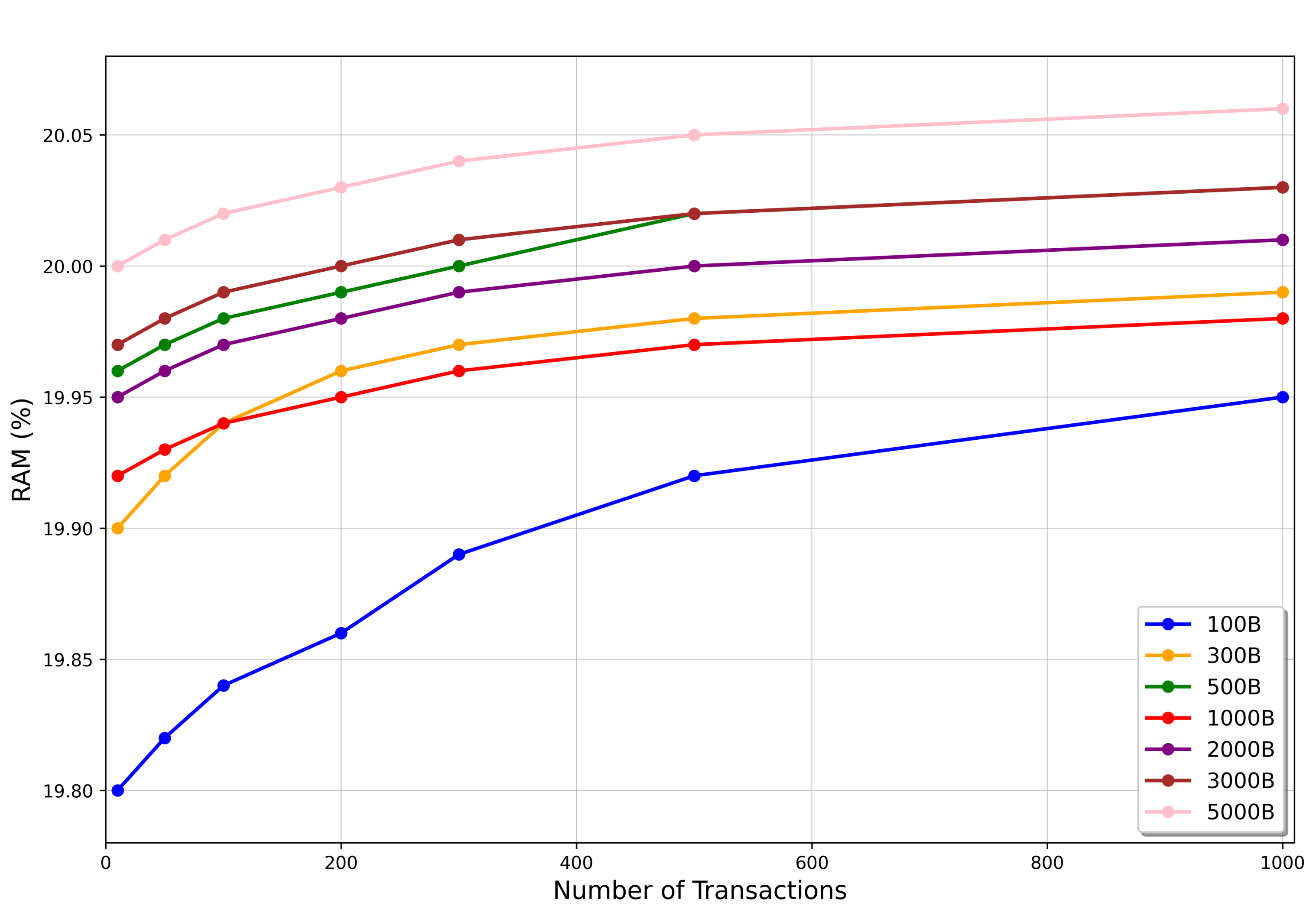}
        \caption{RAM usage without TEE}
        \label{fig:mem_no_tee}
    \end{subfigure}
    \hfill
    \begin{subfigure}{0.50\textwidth}
        \centering
        \includegraphics[width=\linewidth]{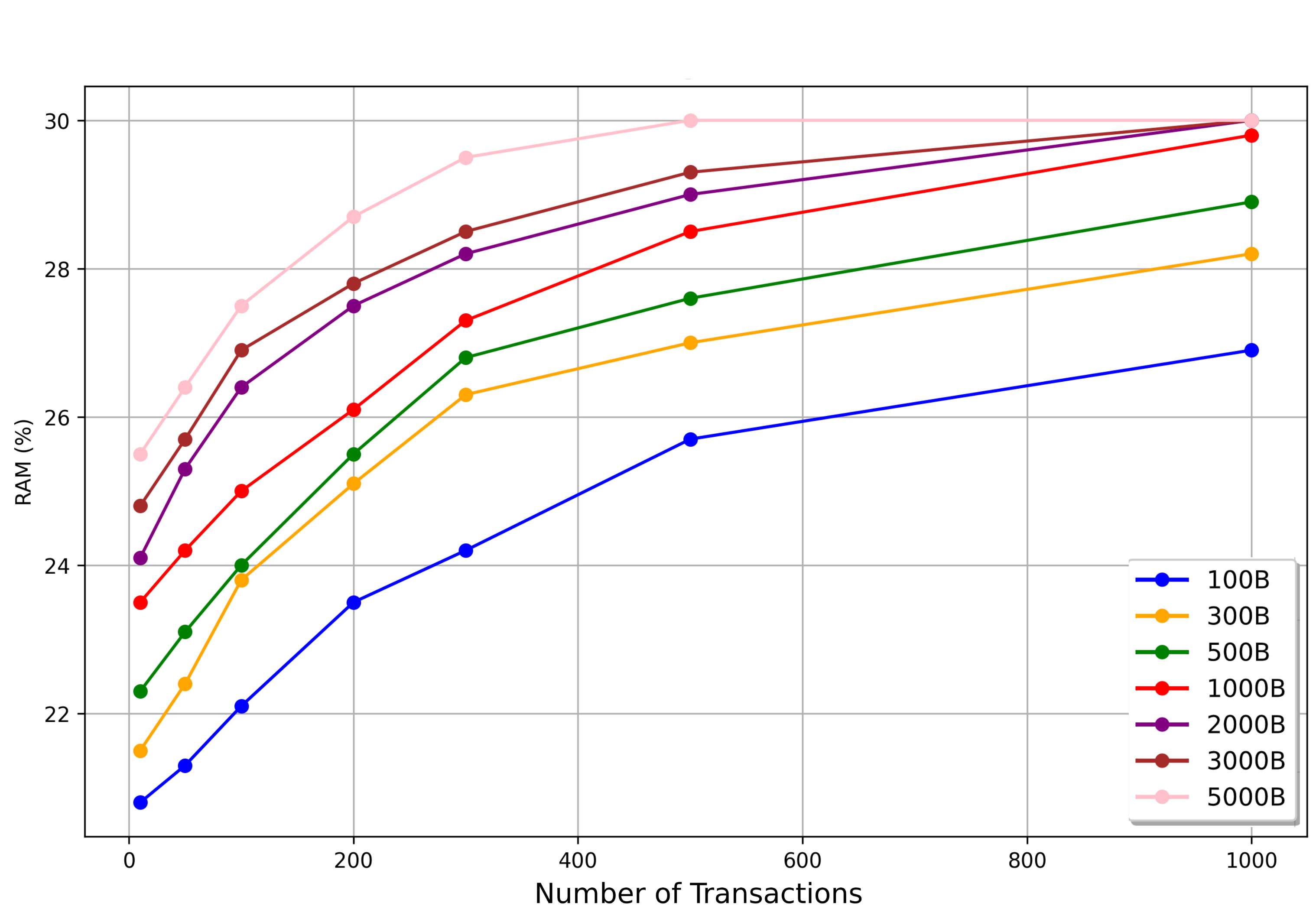}
        \caption{RAM usage with TEE}
        \label{fig:mem_tee}
    \end{subfigure}
    \vspace{-0.5cm}
    \caption{Comparison of RAM consumption when executing the Sequencer (a) outside a TEE and (b) inside a TEE.}
        \label{fig:benchmark_memory}
\end{figure}

\subsection{\textcolor{black}{Security Evaluation}}

\textcolor{black}{In this section, we demonstrate the security guarantees provided by our TEE-secured Sequencer architecture through both theoretical analysis and practical validation of memory protection mechanisms.}

\paragraph{\textcolor{black}{Security Analysis:}} \textcolor{black}{The TEE-secured Sequencer provides by design confidentiality (S1) and integrity (S2) of transaction ordering operations, protecting against privileged attackers who control the host system. Transaction data is protected both inside the TEE enclave during processing (a) and in communications with off-chain components (b).}

\textcolor{black}{\textit{Proof:} To demonstrate these security properties, we analyze protection mechanisms at two levels:
\begin{itemize}
    \item S1.a - Confidentiality inside the TEE: Transaction queues and ordering logic are isolated by hardware-enforced memory encryption. The TEE's secure execution mode prevents unauthorized access to enclave memory, ensuring that pending transactions remain invisible to the host operator.
    \item S2.a - Integrity inside the TEE: The remote attestation mechanism cryptographically verifies both the Sequencer code integrity and runtime metadata (block hash, height, state root). This ensures that transaction processing follows the intended protocol logic without manipulation.
    \item S1.b - End-to-End security: Secure channels established through TEE-terminated TLS protect transaction data during transmission to the Batcher and Proposer components.
    \item  S2.b - Integrity verification: The decentralized attestation mechanism provides on-chain verification that blocks were produced by an authenticated TEE instance, preventing publication of blocks from compromised environments.
\end{itemize}}
 
\textcolor{black}{\paragraph{Security Assessment} We analyze the security guarantees provided by our TEE-secured architecture against the primary attack vectors identified in Section III, demonstrating how the hardware isolation fundamentally alters the attack surface compared to standard Sequencer deployments.}

\textcolor{black}{\textit{Memory Dump Attack:} We conducted practical validation of memory protection by simulating privileged attacker scenarios. The attack procedure involved: (1) disabling the STRICT\_DEVMEM kernel option to enable unrestricted memory access, (2) installing the LiME (Linux Memory Extractor) loadable kernel module to perform physical memory dumps, (3) utilizing binary grep (bgrep) tools to scan memory dumps for transaction patterns, addresses, and MEV opportunities.
Results: Analysis of memory dumps from the TEE-secured Sequencer showed only encrypted data within protected memory pages. Binary analysis tools failed to extract plaintext transaction information, with the TEE's hardware isolation ensuring that sensitive transaction data never appears unencrypted in system memory accessible to privileged attackers. Pattern matching attempts yielded no exploitable transaction details, mempool state, or processing logic information.}

\textcolor{black}{\textit{Transaction Ordering Manipulation:} Traditional Sequencer implementations are vulnerable to sophisticated manipulation techniques including system call interception (ptrace), shared memory modification, signal injection for timing manipulation, and direct memory writes targeting transaction queue structures. These attacks enable reordering of transactions, injection of malicious transactions, and manipulation of processing priorities for MEV extraction.
The TEE's execution model prevents these attack vectors through hardware-enforced isolation mechanisms. For instance, SGX's reverse sandbox architecture treats the external environment (including the OS and hypervisor) as untrusted, creating a protected enclave where transaction processing occurs within encrypted memory pages (EPC - Enclave Page Cache). External attempts to manipulate the transaction ordering process are blocked by the TEE's memory access controls and hardware attestation, maintaining the integrity of the sequencing logic regardless of host system compromise or privileged attacker presence.}

\textcolor{black}{\textit{MEV Extraction Attacks:} Standard deployments expose transaction pools to analysis, enabling attackers to identify profitable opportunities such as DEX arbitrage, sandwich attacks, and liquidation front-running. The visibility of pending transactions allows sophisticated MEV strategies that can systematically extract value from honest users.
Our architecture mitigates these threats through transaction queue isolation within the TEE's protected memory space. The hardware-level encryption (such as SGX's Memory Encryption Engine) ensures that transaction data remains opaque to external observation, even with root privileges. The TEE's controlled entry/exit points (e.g., SGX's ECALL/OCALL mechanisms) prevent unauthorized access to the mempool state, eliminating the information advantage required for MEV extraction. The deterministic and tamper-resistant ordering process ensures that transaction sequencing follows protocol rules rather than attacker preferences, with cryptographic attestation providing verifiable proof of correct execution.}

\textcolor{black}{\paragraph{Security Estimation} The TEE-secured architecture significantly reduces the attack surface compared to standard Sequencer deployments while maintaining the security assumptions of the underlying TEE technology. The decentralized attestation mechanism eliminates single points of trust in the verification process, providing concrete security improvements that justify the performance overhead (7x latency increase, 3x throughput reduction) in scenarios prioritizing transaction ordering integrity.}

\begin{table}[h]
\centering
\renewcommand{\arraystretch}{1.5} 
\resizebox{\columnwidth}{!}{
\begin{tabular}{|l|l|l|l|}
\hline
\textbf{Contract} & \textbf{Gas Usage} & \textbf{\begin{tabular}[c]{@{}c@{}}Gas Fees\\ (GWei)\end{tabular}} & \textbf{Description} \\ \hline
\multicolumn{4}{|c|}{\textit{One-Time Transactions (Automata Contract Deployment)}} \\ \hline
PCCS Router  & 2,352,196  & 13.77 & Manages access to collaterals \\ \hline
DCAP Attestation  & 3,296,655  & 12.21 & The entrypoint contract  to submit a quote \\ \hline
V3 Verifier  & 3,696,655  & 7.56 & Verifies SGX V3 quotes \\ \hline
V4 Verifier  & 4,650,134  & 12.79 & Verifies SGX V4 quotes \\ \hline
PCS DAO  & 2,014,168  & 9.77 & Manages Intel PCS certificates \\ \hline
PCK DAO  & 2,928,849  & 16.22 & Stores PCK keys \\ \hline
FMSPC TCB DAO  & 2,339,367  & 33.11 & Manages TCB status \\ \hline
Enclave ID DAO  & 1,693,126  & 30.48 & Manages SGX enclave IDs \\ \hline
DAO Storage  & 438,565  & 27.93 & Stores attestation data \\ \hline
Verification & 322,25 & 26.12 & Check Sequencer attestation \\ \hline
\multicolumn{1}{|r|}{\textbf{Total}} & \textbf{23,731,965}  & \multicolumn{2}{l|}{\textbf{Average Gas Fees: 18.98}} \\ \hline
\end{tabular}
}
\vspace{0.25cm}
\caption{Breakdown of One-Time Costs for Deployment of Attestation Contracts.}
\label{tab:costs_automata_deployment}
\end{table}

\subsection{Cost Evaluation} 

In this subsection, we analyze the cost of the decentralized attestation mechanism for the TEE-secured Sequencer. Specifically, we evaluate the gas costs associated with the on-chain verification of the attestation quote and explore how these costs vary based on different factors, such as quote size. 

\subsubsection{Breakdown of Attestation Costs}

The attestation process involves several transactions that require computational resources on the blockchain, leading to gas consumption. Based on our implementation using the Automata framework, we categorize the costs into one-time deployment costs (smart contract deployment) and recurring costs (per-attestation verification).

\paragraph{One-Time Deployment Costs} 
Deploying the decentralized attestation system requires multiple smart contracts, each responsible for a specific function. Table \ref{tab:costs_automata_deployment} summarizes the gas usage for deploying key smart contract

The total cost of deploying the decentralized attestation framework is approximately 23.73 million gas units. While this is a one-time cost, it is an important consideration for any entity setting up the attestation system.

\paragraph{Recurring Attestation Costs (Per-Quote Verification)}
Each time the enclave generates a new attestation quote, the system must perform an on-chain verification process. Table \ref{tab:costs_automata_attestation} details the gas usage per attestation. Each attestation transaction consumes approximately 12.55 million gas units, which translates to a significant cost per verification cycle. Given that the enclave periodically generates quotes, the cost per day depends on the frequency of attestations.
    
\begin{table}[h]
\centering
\renewcommand{\arraystretch}{1.5} 
\scalebox{0.75}{
\begin{tabular}{|l|l|l|l|}
\hline
\textbf{Contract} & \textbf{Gas Usage} & \textbf{\begin{tabular}[c]{@{}c@{}}Gas Fees\\ (GWei)\end{tabular}} & \textbf{Description} \\ \hline
\multicolumn{4}{|c|}{\textit{Recurring Transactions (Quote Attestation - 4KB Quote)}} \\ \hline
Verify and Attest Quote  & 8,014,059  & 13.66 & On-chain attestation of an SGX quote \\ \hline
Set Quote Verifier  & 4,544,335  & 16.84 & Registers a new quote verifier on-chain \\ \hline
\multicolumn{1}{|r|}{\textbf{Total}} & \textbf{12,558,394}  & \multicolumn{2}{l|}{\textbf{Average Gas Fees: 15.25}} \\ \hline
\end{tabular}
}
\vspace{0.25cm}
\caption{Breakdown of Recurring Costs for On-Chain Quote Attestation.}
\label{tab:costs_automata_attestation}
\end{table}

\textcolor{black}{The frequency of attestation quote regeneration directly affects both security and operational costs. As described in Section VI.C, our design adopts a hybrid renewal mechanism with a 4-hour validity window, balancing integrity guarantees with gas expenditure. Each renewal currently requires an on-chain verification transaction consuming approximately 12.6 million gas (Table \ref{tab:costs_automata_attestation}), which, assuming a gas price of 15.25 GWei (as observed during our experiments on Ethereum mainnet), translates to ~0.19 ETH per renewal. This cost can be optimized by reducing the attestation quote size or extending the validity window (e.g., 8–12 hours), thereby lowering the number of daily renewals. Furthermore, the verification logic involves only signature and collateral checks, scales linearly with the number of Sequencers, and can be extended with batched submissions to amortize costs. While not negligible, these costs are justified by the security guarantees provided, ensuring verifiable Sequencer integrity without introducing an execution bottleneck on-chain.}

\subsubsection{Impact of Quote Size on Gas Costs}
The size of the attestation quote directly affects gas consumption. Larger quotes require more computational steps for validation, leading to higher gas costs. According to Intel's official documentation, a standard SGX attestation quote has a typical size of approximately 512 bytes, excluding additional metadata and signature overhead \cite{IntelSGXAttestation2024}. However, as said in Section \ref{The_Proposed_solution}, in our implementation, the enclave generates a cryptographic quote that includes both Intel's standard attestation fields (such as MRENCLAVE and TCB status) and additional Sequencer-specific information. Specifically, we extend the quote with fields such as the block hash, block height, state root, timestamp, sequencer nonce, and Prover public key. These additional fields improve the transparency and verifiability of Sequencer operations but also increase the overall quote size. To quantify the impact of these modifications, we simulate gas consumption for different quote sizes, ranging from 512 B to 10 KB. This analysis, shown in, allows us to evaluate the trade-off between quote size growth and attestation verification cost, ensuring that the system remains both secure and cost-efficient. The gas cost for attesting a quote includes the execution of two smart contracts: one for verifying the quote's validity and another for registering the quote verifier. The total gas usage reported in Table \ref{tab:quote_gas_cost} represents the cumulative cost of both operations.

\begin{table}[h]
\centering
\renewcommand{\arraystretch}{1.3}
\scalebox{0.85}{
\begin{tabular}{|c|c|c|}
\hline
\textbf{Quote Size (KB)} & \textbf{Estimated Gas Usage} & \textbf{Estimated Cost (GWei)} \\ \hline
512 B  & 8,636,467  & 14.03 \\ \hline
1 KB  & 9,136,467  & 14.20 \\ \hline
2 KB  & 10,407,443  & 14.15 \\ \hline
4 KB  & 12,690,007  & 15.12 \\ \hline
6 KB  & 13,820,092  & 14.60 \\ \hline
8 KB  & 14,550,541  & 15.50 \\ \hline
10 KB & 15,199,581 & 17.40 \\ \hline
\end{tabular}*}
\vspace{0.2cm}
\caption{Gas Consumption for Varying Quote Sizes}
\label{tab:quote_gas_cost}
\end{table}

The results confirm that gas consumption increases proportionally with quote size. This is expected, as Ethereum's computational model assigns higher costs to larger on-chain data operations.

\subsection{\textcolor{black}{Comparison with Counterparts}}

{\textcolor{black}{We compare our TEE-secured Sequencer approach with established and peer-reviewed rollup schemes, including StarkNet \cite{StarkNet}, Scroll \cite{Scroll}, Optimism, ZkSync Era \cite{zkSync}, and Arbitrum \cite{Arbitrum}. We intentionally exclude experimental or non–peer-reviewed proposals to ensure that the evaluation is based on reliable and widely adopted solutions. Table~\ref{tab:rollup_comparison} summarizes the comparison. L2Beat production data \cite{L2Beat} show notable variations in real-world roll-up performance: Optimism achieves 8–16 TPS, StarkNet operates at 6-10 TPS with occasional spikes to 135 TPS, Scroll maintains 4-12 TPS consistently, ZkSync Era runs at 5–15 TPS with peaks around 60 TPS, and Arbitrum demonstrates 15–35 TPS. In our evaluation, we measured 7–8 TPS on a single-node testbed environment running the full Optimism stack locally, with Layer-1 simulated on Sepolia testnet. These conditions introduced resource constraints, network overhead, and testnet limitations. Despite this setup, our results remain competitive with production rollups.
For transaction fees, StarkNet and Scroll average approximately 0.043 USD and 0.114 USD respectively, while Optimism and Arbitrum achieve lower costs of 0.012 USD and 0.005 USD. Our approach incurs standard Optimism fees plus an additional attestation cost (~\$40–60 per verification cycle), which, when amortized over transaction batches, represents a security overhead aimed at mitigating operator manipulation and MEV extraction. Latency-wise, our system shows confirmation times of 21–25 seconds, compared to 1–10 seconds for production rollups.}

\begin{table}[h]
\centering
\small
\scalebox{0.83}{
\begin{tabular}{|l|c|c|c|l|l|}
\hline
\textbf{Rollup} & \textbf{TPS} & \textbf{Lat.} & \textbf{Fee} & \textbf{Security} & \textbf{Arch.} \\
\hline
Optimism & 8-16 & 1-10s & \$0.012 & Centralized Seq. & Single Seq. \\
\hline
StarkNet & 2-8 & 10-30s & \$0.043 & ZK Proofs & Centralized \\
\hline
Scroll & 2-4 & 1-3s & \$0.114 & ZK Proofs & Centralized \\
\hline
ZkSync & 5-15 & 2s & \$0.002 & ZK Proofs & Centralized \\
\hline
Arbitrum & 15-35 & 1-2s & \$0.005 & Fraud Proofs & Centralized \\
\hline
\textbf{Ours} & \textbf{7-8} & \textbf{21-25s} & \textbf{\$0.012*} & \textbf{TEE + Dec. Att.} & \textbf{Enhanced Seq.} \\
\hline
\end{tabular}}
\caption{Comparison of rollup solutions (*+\$40-60/attestation).}
\label{tab:rollup_comparison}
\end{table}

\section{Conclusion}
\label{Conclusion}

Our paper proposed a security solution for Rollup's Sequencers using Trusted Execution Environments and a decentralized attestation mechanism. The approach ensures both the integrity and confidentiality of the Sequencer’s execution, while at the same time ensures the decentralization of the integrity verification process through smart contracts. This shift not only mitigates the risk of a single point of failure but also aligns with the core principles of blockchain, promoting transparency and distributed trust. 

We experimentally evaluated our approach by comparing a TEE-secured Sequencer with its unprotected counterpart, focusing on key performance metrics such as startup time, latency, throughput, CPU, and memory consumption. The results confirmed that while executing the Sequencer within a TEE introduces measurable overhead — increasing latency approximately from 2.5–3.5 seconds to over 25 seconds and reducing throughput from about 25 TPS to around 7-8 TPS — the proposed solution remains suitable for scenarios prioritizing security and integrity over transaction speed. 
It is worth noting that the performance measurements are based on Intel SGX, which is known to introduce higher overhead due to its smaller and more isolated Trusted Computing Base. Replacing SGX with TDX in future implementations would likely improve performance while maintaining similar attestation guarantees, as TDX has a broader execution model with lower resource constraints.
Furthermore, the cost analysis of the decentralized attestation mechanism revealed that while deploying and managing this decentralized infrastructure incurs significant initial gas costs, recurring costs per attestation can be optimized by careful management of quote size and attestation frequency.

\section*{Acknowledgment}
This project has received funding from the European Union's Horizon Europe Research and Innovation Programme under Grant Agreement No. 101070670 (ENCRYPT - A Scalable and Practical Privacy-preserving Framework). 
\\The work made in this paper was also funded by the European Union under NextGenerationEU. PRIN 2022 Prot. n.  202297YF75.
\\We further acknowledge the valuable financial support of Trillion.xyz, which contributed to the realization of this research.

\bibliographystyle{IEEEtran}
\bibliography{reference}

\vspace{-20pt}

\begin{IEEEbiography}[{\includegraphics[width=0.8in,height=1.25in,clip,keepaspectratio]{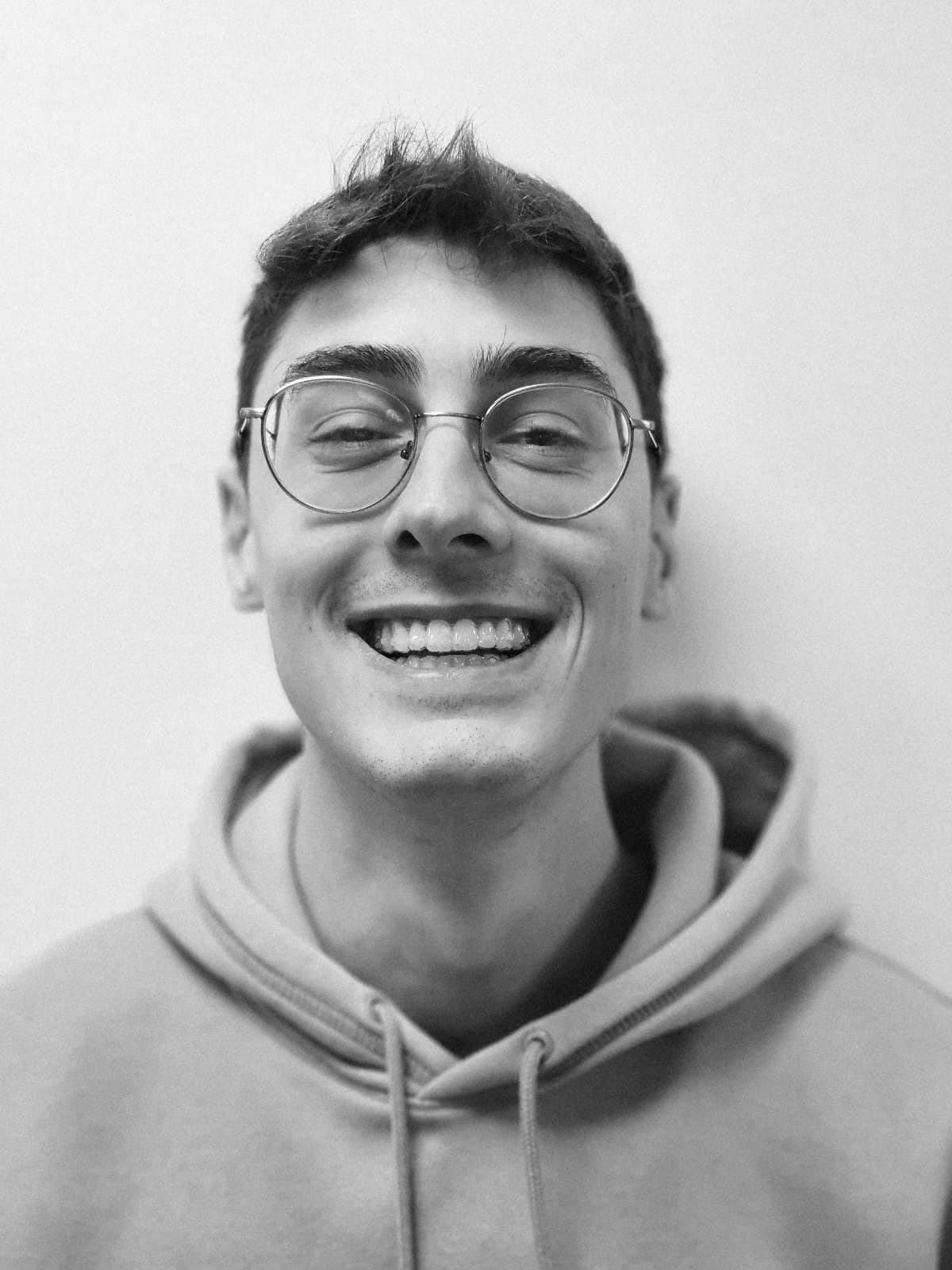}}]{Giovanni Maria Cristiano} \scriptsize is a PhD student at the University of Naples "Parthenope". His research interests include federated learning, privacy-preserving machine learning, and secure distributed systems. He is involved in European and National research projects focused on decentralized learning and the protection of sensitive data in collaborative environments.

\end{IEEEbiography}

\vspace{-10pt}

\begin{IEEEbiography}[{\includegraphics[width=1in,height=1.25in,clip,keepaspectratio]{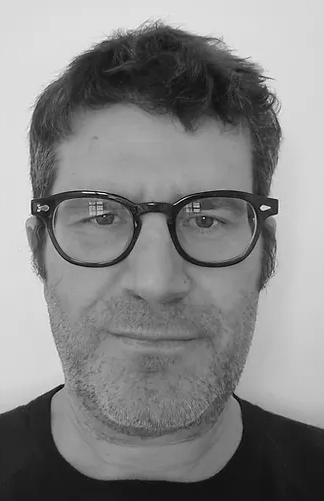}}]{Salvatore D'Antonio}
\scriptsize is an Associate Professor at the University of Naples 'Parthenope'. He is an expert in network monitoring, network security and critical infrastructure protection. He was the Coordinator of two EU research projects on critical infrastructure protection, namely INSPIRE and INSPIRE-INCO.
\end{IEEEbiography}

 \begin{IEEEbiography}[{\includegraphics[width=0.8in,height=1.25in,clip,keepaspectratio]{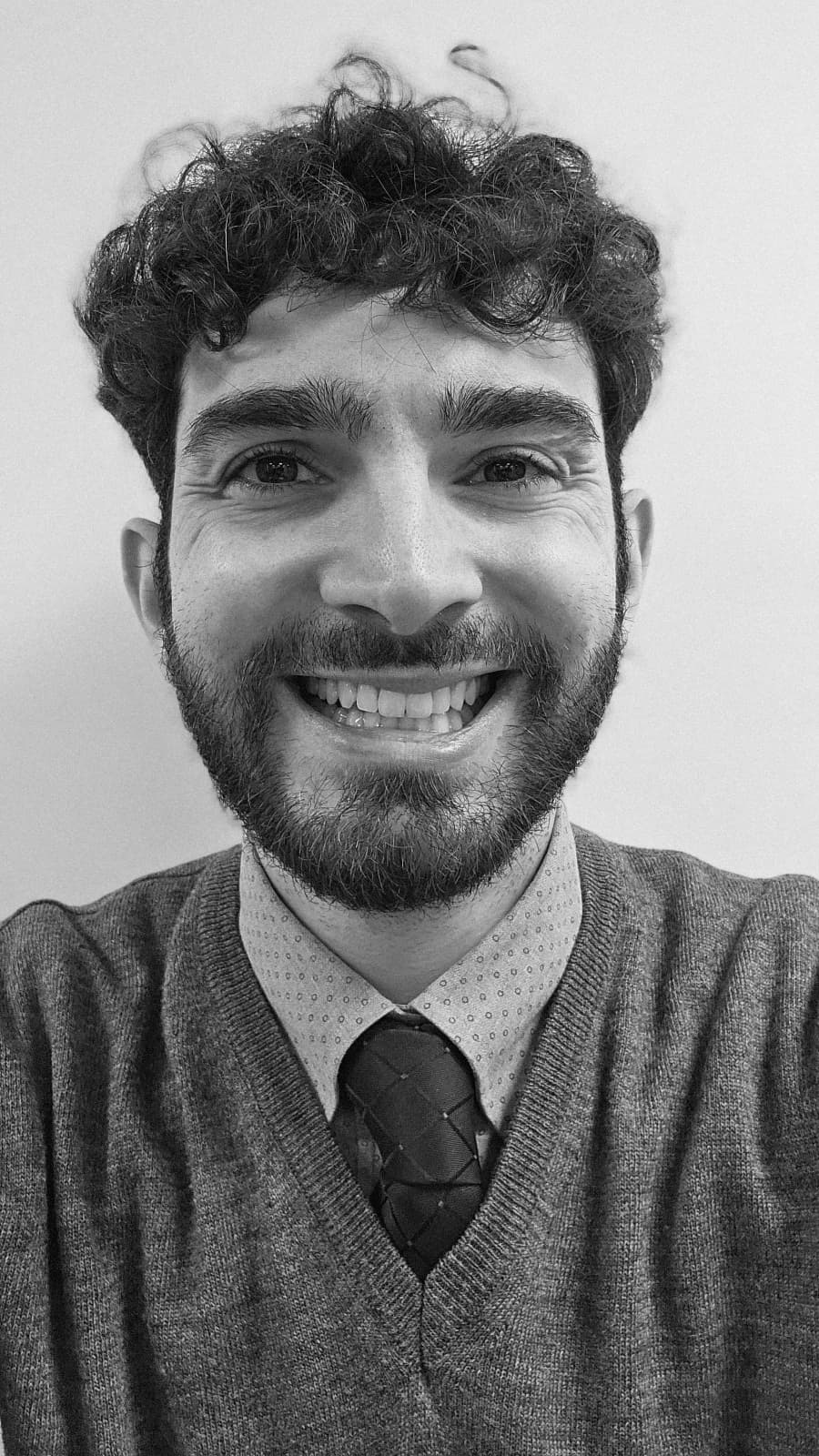}}]{Jonah Giglio} \scriptsize is a PhD student at the University of Naples "Parthenope". His research focuses on the security and dependability of computer systems, with a particular emphasis on hardware-assisted trusted computing and secure architectures. He is involved in national and European research projects on confidential computing and computing continuum, contributing to the design and integration of secure system components.
\end{IEEEbiography}

\begin{IEEEbiography}[{\includegraphics[width=1in,height=1.25in,clip,keepaspectratio]{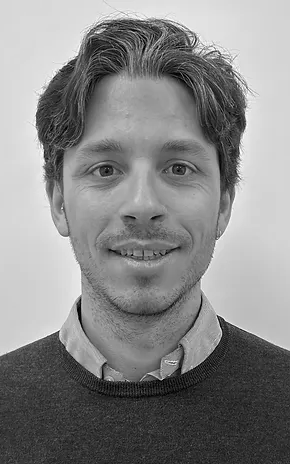}}]{Giovanni Mazzeo} \scriptsize PhD,
is an Associate Professor at the University of Naples 'Parthenope'. His research field is the security and dependability of computer systems, with a particular focus on hardware-assisted trusted computing. He was principal investigator of several Italian and European research projects in the context of confidential computing. He has covered key technical roles such as leading work packages on the definition of the architecture and of the integration process. 
\end{IEEEbiography}

\begin{IEEEbiography}[{\includegraphics[width=0.8in,height=1.25in,clip,keepaspectratio]{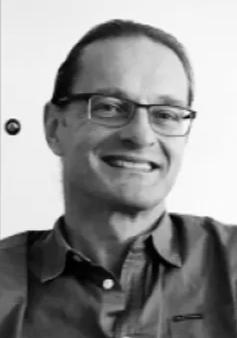}}]{Luigi Romano} \scriptsize PhD, is a Full Professor at the University of Naples 'Parthenope'. His research interests are system security and dependability, with focus on Critical Infrastructure Protection. He has worked extensively as a consultant for industry leaders in the field of security- and safety-critical computer systems. He was one of the members of the ENISA expert group on Priorities of Research On Current and Emerging Network Technologies (PROCENT). 
\end{IEEEbiography}

\end{document}